\documentclass[aps,pra,reprint,superscriptaddress,amssymb,amsmath]{revtex4-1}
  \usepackage[utf8]{inputenc}
  \usepackage[T1]{fontenc}
  \usepackage[english]{babel}
  \usepackage{times}
  \usepackage{amssymb,amsmath,amsfonts,amsthm}
  \usepackage{mathtools}
  \usepackage{verbatim}
  \usepackage{enumerate}
  \usepackage{graphicx}
  \usepackage{hyperref}
  \usepackage{bbm}
  \usepackage{booktabs}
  
  \usepackage[caption=false]{subfig}
  \usepackage{mathrsfs}
  \usepackage{color}
  \usepackage{bm}
  \usepackage{natbib}
\usepackage{graphicx}
\usepackage{multirow}
\usepackage{bm}

\begin{document}                                                 
\title{Experimental free-space quantum secure direct communication and its security analysis}

\author{Dong Pan}
\thanks{These authors have contributed equally to this work}
\affiliation{State Key Laboratory of Low-dimensional Quantum Physics and Department of Physics, Tsinghua University, Beijing 100084, China}
\affiliation{Frontier Science Center for Quantum Information, Beijing 100084, China}

\author{Zaisheng Lin}
\thanks{These authors have contributed equally to this work}
\affiliation{Frontier Science Center for Quantum Information, Beijing 100084, China}
\affiliation{School of Information Science and Technology, Tsinghua University, Beijing 100084, China}
\affiliation{Beijing National Research Center for Information Science and Technology, Beijing 100084, China}
\affiliation{Beijing Academy of Quantum Information Sciences, Beijing 100193, China}

\author{Jiawei Wu}
\affiliation{State Key Laboratory of Low-dimensional Quantum Physics and Department of Physics, Tsinghua University, Beijing 100084, China}
\affiliation{Frontier Science Center for Quantum Information, Beijing 100084, China}

\author{Haoran Zhang}
\affiliation{State Key Laboratory of Low-dimensional Quantum Physics and Department of Physics, Tsinghua University, Beijing 100084, China}
\affiliation{Frontier Science Center for Quantum Information, Beijing 100084, China}

\author{Zhen Sun}
\affiliation{Frontier Science Center for Quantum Information, Beijing 100084, China}
\affiliation{School of Information Science and Technology, Tsinghua University, Beijing 100084, China}

\author{Dong Ruan}
\affiliation{State Key Laboratory of Low-dimensional Quantum Physics and Department of Physics, Tsinghua University, Beijing 100084, China}
\affiliation{Frontier Science Center for Quantum Information, Beijing 100084, China}

\author{Liuguo Yin}
\email{yinlg@tsinghua.edu.cn}
\affiliation{Frontier Science Center for Quantum Information, Beijing 100084, China}
\affiliation{School of Information Science and Technology, Tsinghua University, Beijing 100084, China}
\affiliation{Beijing National Research Center for Information Science and Technology, Beijing 100084, China}
\affiliation{Beijing Academy of Quantum Information Sciences, Beijing 100193, China}

\author{Guilu Long}
\email{gllong@tsinghua.edu.cn}
\affiliation{State Key Laboratory of Low-dimensional Quantum Physics and Department of Physics, Tsinghua University, Beijing 100084, China}
\affiliation{Frontier Science Center for Quantum Information, Beijing 100084, China}
\affiliation{School of Information Science and Technology, Tsinghua University, Beijing 100084, China}
\affiliation{Beijing National Research Center for Information Science and Technology, Beijing 100084, China}
\affiliation{Beijing Academy of Quantum Information Sciences, Beijing 100193, China}

\date{\today}

\begin{abstract}
We report an experimental implementation of free-space quantum secure direct communication based on single photons. The quantum communication scheme uses phase encoding, and the asymmetric Mach-Zehnder interferometer is optimized so as to automatically compensate phase drift of the photons during their transitions over the free-space medium. At the 16 MHz pulse repetition frequency, an information transmission rate of 500 bps over a 10 meter free space with a mean quantum bit error rate of 0.49\%$\pm$0.27\% is achieved. The security is analyzed under the scenario that Eve performs the collective attack for single-photon state and the photon number splitting attack for multi-photon state in the depolarizing channel. Our results show that quantum secure direct communication is feasible in free space.
\end{abstract}

\maketitle
\section{Introduction}
Information security and data encryption~\cite{sun2014smart,cai2017visible} have risen to a pivotal position in the digital information era. The development of quantum communication provides us with new approaches for secure communication tasks, with the benefit of provable security provided by quantum mechanical laws. Quantum key distribution (QKD) protocol was proposed by Bennett and Brassard in 1984 (called BB84 QKD protocol)~\cite{bennett1984quantum} to perform key exchange between legitimate distant users. Hitherto, QKD has been well developed in optical fiber, laying foundation for the establishment of quantum communication networks~\cite{wang2010field,sasaki2011field,wang2014field}. Compared with the fiber, the free-space channel is also considered as a befitting link for quantum communication. The atmosphere has several high transmission windows at particular wavelengths, which allows low-loss light transmission. Quantum communication can be established by using a free-space channel~\cite{liu2020drone} for rough areas where optical fiber networks are not constructed. In addition, free-space quantum communication is valuable for long-distance quantum communication, combining earth-to-satellite and satellite to satellite communications. Due to nonbirefringence for the propagation of light in atmosphere, the polarization of single photon is maintained well, most free-space quantum communications are implemented using polarization encoding \cite{buttler1998practical,bienfang2004quantum,schmitt2007experimental,tannous2019demonstration,liao2017satellite}. QKD ensures security through detection of eavesdropping on-site. Therefore QKD transmits random numbers first, and if it can assure no eavesdropping, the random numbers are adopted as keys for use to encrypt the message in a subsequent classical communication. But it cannot prevent the eavesdropper from obtaining the transmitted ciphertext.

In the past two decades, quantum secure direct communication (QSDC) was proposed and developed \cite{long2002theoretically,deng2003two,deng2004secure,wang2005quantum}. QSDC directly conveys safely secret messages over quantum channel. Demonstration experiments have contributed the key technologies of QSDC, such as frequency coding \cite{hu2016experimental}, quantum memory \cite{zhang2017quantum}, fiber entanglement source \cite{zhu2017experimental}, and practical system for intra-city applications \cite{qi2019implementation}. Up to now, this philosophy has been extended to numerous different theoretical proposals aimed to directly convey secret information over quantum channel, which guarantees security by ensuring eavesdropper cannot simultaneously access to the two parts of a correlated quantum state \cite{long2002theoretically,deng2003two,wang2005quantum,marino2006deterministic,shapiro2014secure} or by encrypting information with quantum state \cite{deng2004secure,pirandola2008quantum,pirandola2009confidential,lum2016quantum}. Recently, the measurement-device-independent (MDI) theories of QSDC have been established \cite{zhou2020measurement,niu2018measurement,gao2019long}, MDI scheme for the single photon-based QSDC was given in Ref. \cite{zhou2020measurement}, and that for the entanglement-based QSDC protocols in Refs. \cite{long2002theoretically,deng2003two} is provided in Ref. \cite{niu2018measurement}. The scheme that is secure against all defects in devices in QSDC, namely, the device-independent QSDC, was given in Ref. \cite{zhou2019device}.

Against the aforementioned background, our main contributions are as follows. First, to the best of our knowledge, we report the first fully operational system for free-space QSDC with phase encoding. The transmitter and receiver modules are further developed by utilizing the most common fiber optical components. A round-trip optical architecture can also mitigate the problem of phase drift in the free-space channel so as to realize a stable QSDC. Second, the security of the QSDC system is analyzed under the photon number splitting (PNS) attack for multi-photon components. The Gottesman-Lo-L{\"u}tkenhaus-Preskill (GLLP) theory \cite{gottesman2004security} and decoy state \cite{hwang2003quantum,wang2005beating,lo2005decoy} can be extended into our model to analyse the security. One surprising result is that we can achieve secure information transmission by the two-photon component, which is consistent with the results of two-way QKD \cite{deng2004bidirectional,lucamarini2005secure,lu2019ambiguous}, a special case of the DL04 QSDC protocol \cite{deng2004secure}. This paper is arranged as follows. In Section \ref{sec:Expim}, we review the details of the single photon-based QSDC protocol and show how we run it on a free-space experimental system with phase encoding. In Section \ref{sec:Experimental results}, we present the experimental results. In Section \ref{sec:Security analysis}, we analyze the security of the QSDC system. Finally, conclusions are given in Section \ref{sec:Conclusions}.

\section{Experimental implementation}
\label{sec:Expim}
\subsection{Protocol}
\label{subsec:Protocol}
The DL04 QSDC protocol \cite{deng2004secure} realized in this work has the following steps.

(1) Bob randomly chooses either the basis $Z$ or $X$ for preparing a sequence of single photons, which are subsequently transmitted to Alice. Each of the photons is in one of four quantum states $\{\vert0\rangle, \vert1\rangle, \vert+\rangle=(\vert0\rangle+\vert1\rangle)/\sqrt{2}, \vert-\rangle=(\vert0\rangle-\vert1\rangle)/\sqrt{2}\}$. One could implement this random selection using a quantum random number generator \cite{zhou2019practical}.

(2) After receiving the photons from Bob, Alice randomly chooses some photons as samples for detecting eavesdropping. For these selected photons, Alice measures each of them by using either the basis $Z$ or $X$ randomly and then announces the positions of the sample together with the measurement basis and outcomes. Alice and Bob obtained the detection bit error rate (DBER) through a classical authenticated channel. 

(3) If the DBER is lower than a predetermined threshold, the information encoding process continues. Alice performs $I=\vert0\rangle\langle0\vert+\vert1\rangle\langle1\vert$ or $Y=i\sigma_{y}=\vert0\rangle\langle1\vert-\vert1\rangle\langle0\vert$ on the remaining photons to encode the secret information bit 0 or 1 and then returns them to Bob. She will also encode some photons randomly for error-checking. Otherwise, the communication process is aborted. 

(4) After receiving the photon sequence, Bob deterministically decodes the secret information. Bob obtains a quantum bit error rate (QBER) by discussing with Alice on the checking bits.

There are two error rates in the DL04 QSDC protocol, the DBER and the QBER, which ensure the security of the first transmission and the reliability of the second transmission, respectively.

\subsection{Phase encoding}
\label{subsec:Phase encoding}
\begin{figure*}[htbp]
\centering\includegraphics[width=17.5cm]{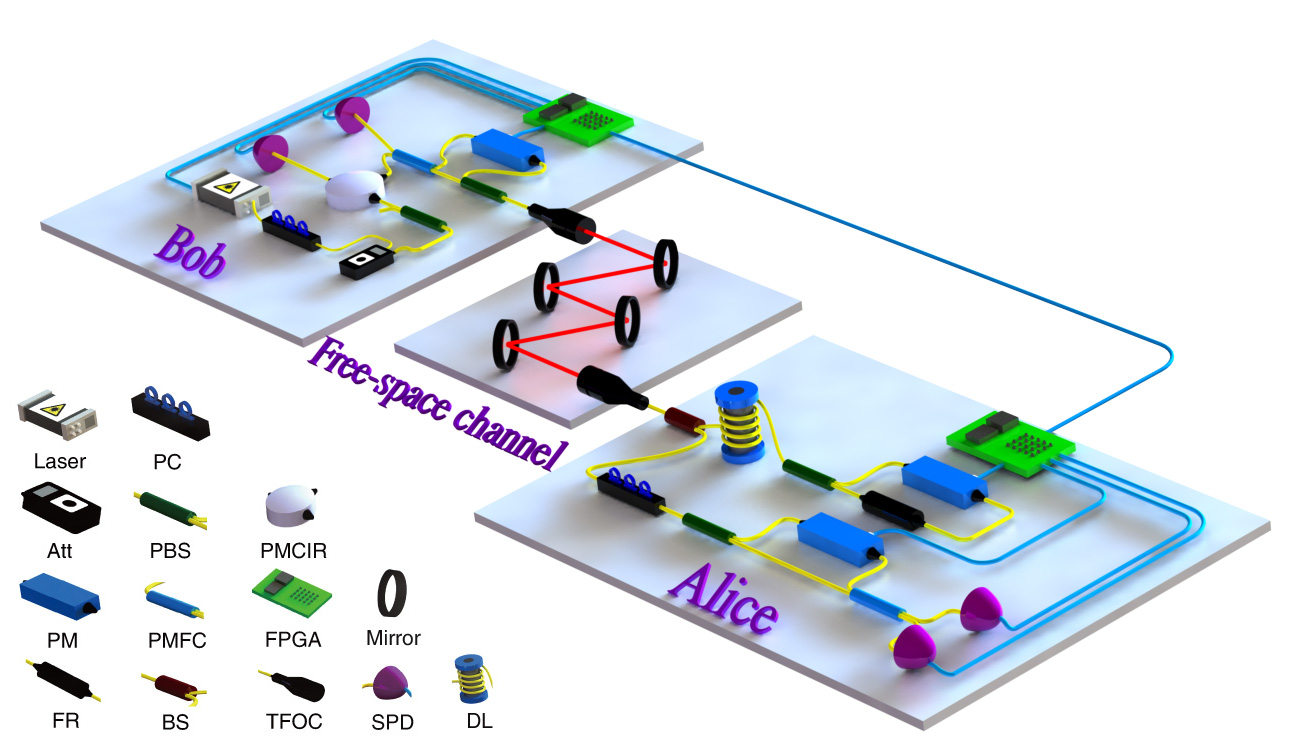}
\caption{Schematic diagram of free-space QSDC system. Att, attenuator; BS, beamsplitter; DL, delay line; FPGA, field-programmable gate array; FR, Faraday rotator; PBS, polarization beam splitter; PC, polarization controller; PM, phase modulator; PMCIR, polarization maintaining circulator; PMFC, polarization maintaining fiber coupler; SPD, single-photon detector; TFOC, triplet fiber optic collimator. Blue, yellow, and red lines are the electric line, optical fiber line, and free space path, respectively.}
\label{Fig:1}
\end{figure*}

The schematic of our experimental setup is shown in Fig. \ref{Fig:1}. The system is comprised of two legitimate users' optical setups and a free-space channel between them. The apparatus of Alice and Bob all adopt fiber-optic components. Some low-absorption atmospheric spectral windows in the near-infrared, such as regions of $\lambda$\textasciitilde850 nm and $\lambda$\textasciitilde1550 nm are usually considered for free-space quantum communications. Our system works at a wavelength of 1550 nm to take advantage of a peak in the typical atmospheric transmission window and the low attenuation dip in fiber-optic components. 

The laser pulses are emitted at Bob with a repetition frequency of 16 MHz and a pulse width of 200 ps. They are reduced to a specific attenuated level at the input of Bob's station. To be more specific, Bob modulates a random phase $\phi_{1}^{B}\in\{0, \pi/2, \pi, 3\pi/2\}$ on the pulse by using his phase modulator (PM) located in the long-path of the asymmetric Mach-Zehnder interferometer. It is equivalent to the preparation of four initial states in the DL04 QSDC protocol. The photons are transported to a triplet fiber optic collimator (TFOC) where they are output to a free-space channel and then collected by Alice's collimator for coupling into the single-mode fiber. In our proof-of-principle experimental demonstration, Alice's and Bob's collimators are separated by 10 m with four mirror reflections. A 50/50 beam splitter (BS) in Alice's system randomly reflects or transmits the incoming photons to two different paths: the lower and upper path in Fig. \ref{Fig:1}, one for detecting eavesdropping and the other for encoding secret information. For the lower path, Alice detects the photon with her interferometer by randomly applying phase modulation $\phi_{1}^{A}\in\{0, \pi/2\}$ to the pulse passing over the long-path, then a DBER is obtained by public discussion between Alice and Bob. By contrast, in the upper path, an encoding operation $I$ or $Y$ is performed on the pulse (previously passing over long-path at Bob) by adding a phase $\phi_{2}^{A}=0$ or $\phi_{2}^{A}=\pi$ after it passes through the Faraday rotator (FR).  Finally, by the time of the pulse arrives back to Bob's station, Bob applies phase modulation $\phi_{2}^{B}$ to the pulse for finishing measurement according to the initial phase modulation that he has imposed. To estimate the QBER, the measurement results of checking bits are compared with Alice's encoding. The photons are detected by InGaAs avalanche photodiodes gated in Geiger mode and cooled to -50 $^{\circ}$C, with a gate width of 1 ns and an efficiency of 5.57\% as well as a dark count probability of $1\times10^{-6}$ per gate. 

In this setup, all pulses propagate over a loop with the FR and the PM to realize information encoding. The Faraday mirror in Muller's scheme \cite{muller1997plug} is replaced by the FR. All pulses only pass through the PM once compared with the Faraday mirror as a reflection terminal, so this loop has less attenuation than the original Muller's scheme. It will help to improve the repetition rate of our QSDC system. The pulses are delivered through the same optical path to convey information, the phase is very stable, and the light propagation with an FR automatically compensates for all polarization fluctuations in the optical links. Furthermore, this system has a low requirement on PM, since the PM is consistent with the conventional one that only requires both its input, and output fibers are polarization-maintaining fibers \cite{wang2018practical}. The polarization controller (PC) located at Alice site is used to compensate polarization drift in the fiber so that the pulses are completely transmitted at the polarization beam splitter (PBS), guiding the short (long) path pulse which comes from Bob into Alice's long (short) path. This free-space QSDC system is controlled as well as synchronized by two field-programmable gate array (FPGA) devices, and specific computer software programs are developed at Alice's and Bob's terminal.

\section{Experimental results}
\label{sec:Experimental results}
\begin{figure}[htbp]
\centering
\includegraphics[width=8.5cm]{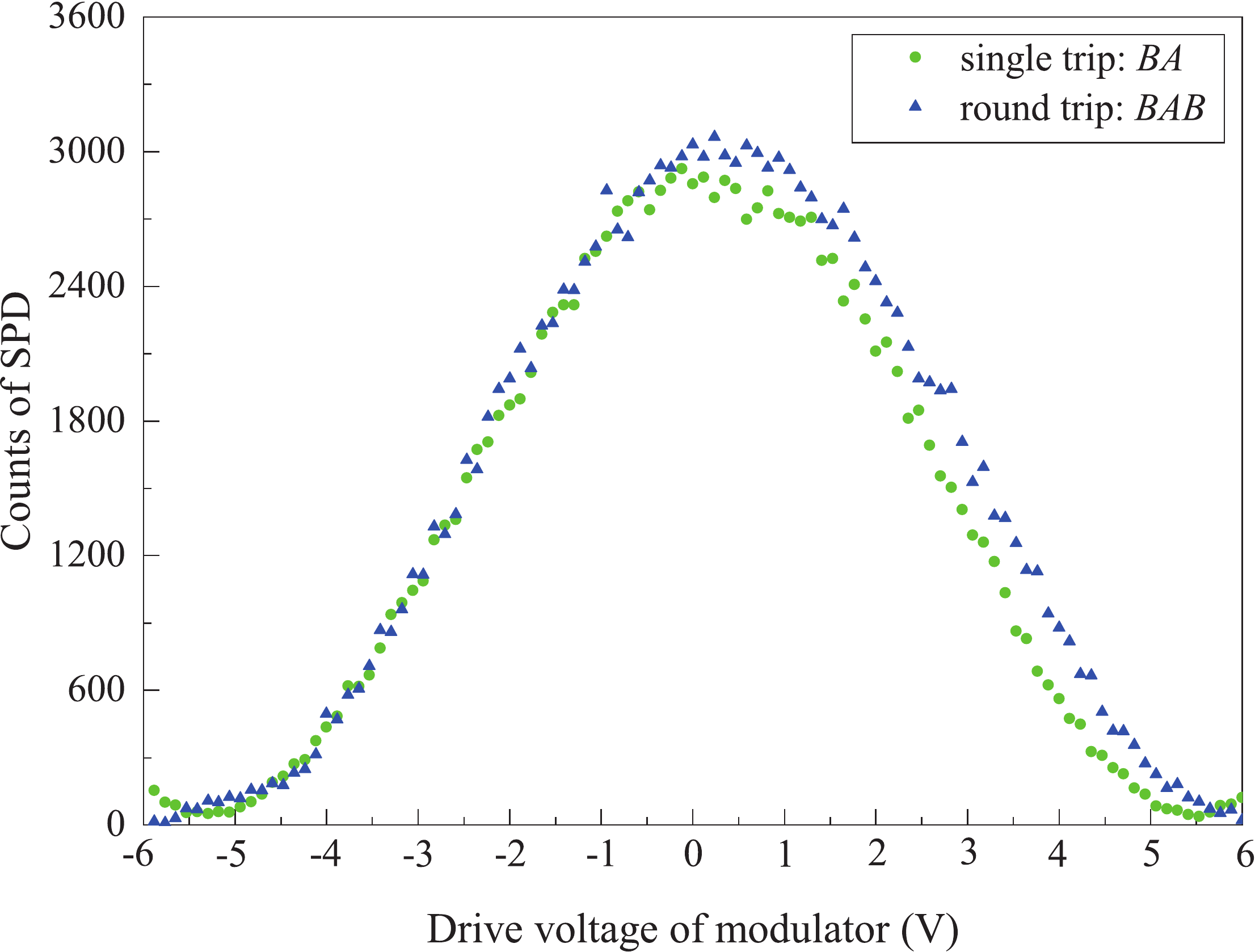}
\caption{Interference fringes. Driving voltage from -6 V to +6 V with a half-wave voltage of 4.8 V and a step of about 0.1 V. The interference fringe of single-trip (photons transmitted from Bob-to-Alice) is obtained from Alice's detection. More specifically, the counts are recorded by Alice's SPD at each step when she drives the voltage of her PM. By contrast, when the photons received by Bob (after their trip Bob-Alice-Bob), he drives the voltage of his PM and records counts by his SPD to obtain the interference fringe of round-trip.}
\label{Fig:2}
\end{figure}

The experiment is conducted in a lab platform. Figure \ref{Fig:2} shows the interference fringes. Both curves are coincident with a sinusoidal pattern. Interference visibility of single trip (Bob-to-Alice) and round trip (Bob-to-Alice-to-Bob) is 97.37\% and 99.48\%, respectively. Although the light is susceptible to scatter in free-space, producing phase aberrations which perturb quantum bits, stable interference can still be observed in our experiment system. 

\begin{figure}[htbp]
\centering
\includegraphics[width=8.5cm]{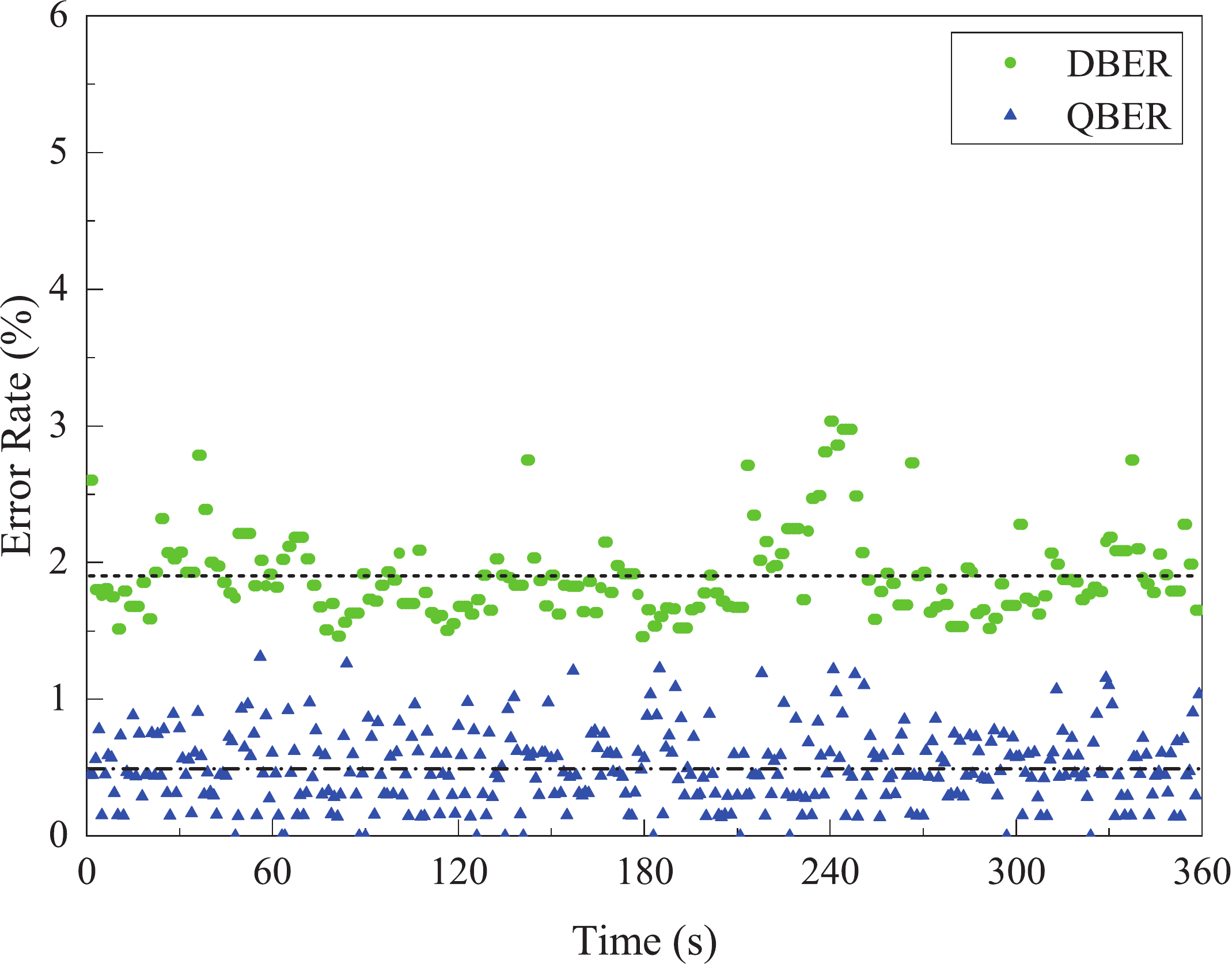}
\caption{Error rates during image file transmission. Dashed lines represent the mean values of DBER and dash-dotted shows the mean values of QBER. The definition of DBER and QBER is given in Section~\ref{subsec:Protocol}, while the experimental approach for accessing them is introduced in Section~\ref{subsec:Phase encoding}.}
\label{Fig:3}
\end{figure}

To guarantee the reliable transmission of secret information, low-density parity-check code \cite{thangaraj2007applications,qi2019implementation} is applied to our free-space QSDC system, and the compensation algorithm that aims to eliminate phase shift of single photon in the free-space channel is equipped. A transmission rate of 500 bps is obtained, consequently, files of reasonable sizes, such as text, picture, and audio, can be transmitted directly over the quantum channel by running our system. In the experiment test, Alice transmits an image of size 800$\times$525 pixels (194 k) to Bob, and Fig. \ref{Fig:3} shows the variation of DBER and QBER during the transmission time. The average of DBER and QBER during image transmission is 1.90\%$\pm$0.32\% and 0.49\%$\pm$0.27\%, respectively. High visibility of the interferometer is crucial to obtain a low error rate in our free-space QSDC system. The QBER through a round-trip optical path is obtained where phase drifts are auto-compensated by the modified Muller's scheme, while the DBER is detected through a single-trip optical path using phase compensation algorithm to mitigate phase shifts. This active compensation is not as efficient as the auto-compensation, therefore DBER is higher than QBER, as shown in Fig. \ref{Fig:3}. The phenomenon of DBER higher than QBER is consistent with the result of Fig. \ref{Fig:2}, in which the interference visibility of the single-trip is lower than that of the round-trip, since the interference visibility has an important influence on the bit error rate of phase-encoding scheme. The round-trip interferometer which has the same optical path for the two interfering pulses produces better interference than the single-trip where the two interfering pulses have only approximately the same optical path. Therefore, the interference visibility of round-trip is higher than that of single-trip. The interference visibility test is generally used to assess the performance of interferometer while the effect of the dark count of single-photon detectors is included. We maintain the detectors' maximum count of $\sim$3000 by improving the light intensity during the interference visibility test of round trip. In this count rate, the influence of the dark count could be ignored. As a result, the count curves are given as Fig. \ref{Fig:2}.

\section{Security analysis}
\label{sec:Security analysis}
The secrecy capacity lower bound of the DL04 QSDC has given in Ref. \cite{qi2019implementation} according to the Wyner's wiretap channel theory \cite{wyner1975wire}, which can be written as
\begin{equation}
C_{s}=\max \limits_{\{p_{0}\}}\left\{I(A:B)-I(A:E)\right\},
\end{equation}
where $I(A\!\!:\!\!B)$ is the mutual information between Alice and Bob, while $I(A\!\!:\!\!E)$ is the maximum information that Eve can steal, and $p_{0}$ is the probability Alice performs operation $I$ during her information encoding. Hence, $C_{s}$ defines the asymptotic information rate at which Alice can convey to Bob over the quantum channel with the guarantee that Eve has negligible information about the transmitted secret information. Remarkably, the asymptotic regime cannot be met for practical implementation, which has been fully considered in QKD \cite{tomamichel2012tight}. The finite size of a block in the practical implementation of block-transmission-based QSDC \cite{long2002theoretically,deng2003two,deng2004secure} is actually the finite-size regime, and the block size would affect the security of QSDC. However, the finite-size analysis of QKD cannot be directly invoked for QSDC, since negotiating random secret key bits is different from transmitting secret information bits. The finite-size effect of QSDC would be an interesting direction for future research.

\subsection{Photon number splitting attack}
\label{subsec:Photon number splitting collective attack}
The general collective attacks on single photon have been taken into account in many works \cite{lu2011unconditional,henao2015practical,qi2019implementation,wu2019security}. However, practical quantum communication systems are usually implemented with weak coherent light sources. The pulse generated from such a light source can be written as a mixture of Fock states $\rho=\int(1/2\pi)d\theta|\sqrt{\mu}e^{i\theta}\rangle\langle\sqrt{\mu}e^{i\theta}|=\sum_{n}p(n,\mu)\vert n\rangle \langle n\vert$, in which the number $n$ of photons follows the Poisson distribution $p(n,\mu)=e^{-\mu}\mu^{n}/n!$ with mean photon number $\mu$ and phase $\theta$. It occasionally emits multiple photons. Unfortunately, the pulses containing multiple photons cannot be secure in some quantum communication protocols when they are under the PNS attack \cite{hwang2003quantum}, namely, Eve splits one of the photons from the pulse that contains two or more photons for measuring. Here, we suggest a photon number splitting attack according to the two-way characteristic of the DL04 QSDC, which combines the PNS attack as well as the collective attack. Hence, the security analysis of this system is given in the context of both the general collective attack on single photon and the PNS attack on multiple photons.

\begin{figure}[htbp]
\centering\includegraphics[width=8.5cm]{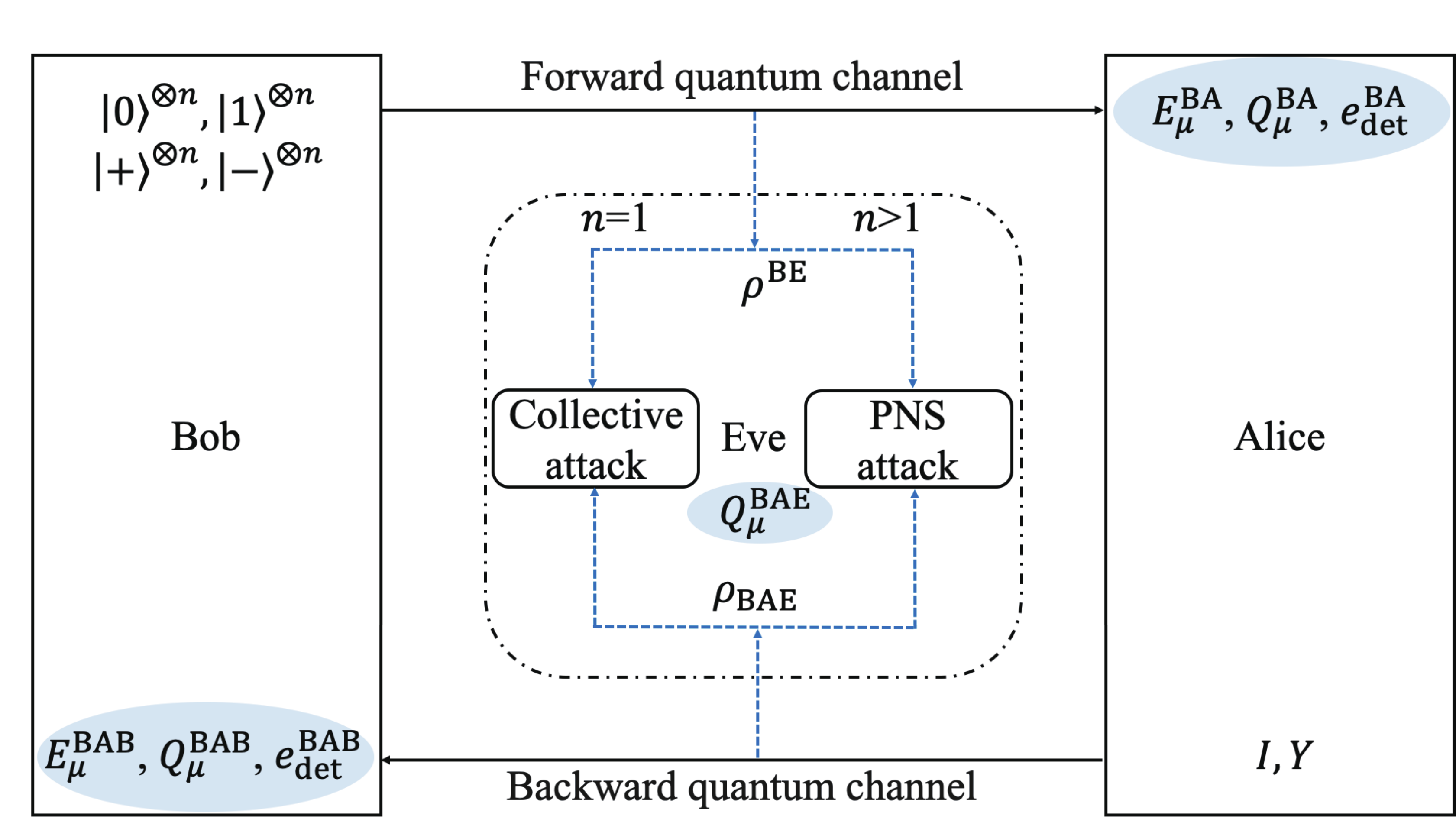}
\caption{The illustration of Eve's attack strategies. $n$, the number of photons in a pulse in the forward quantum channel; $E_{\mu}^{\rm BA}$ is the error rate of the Bob-Alice channel, which is also called as DBER; $Q_{\mu}^{\rm BA}$, the overall signal gain of Alice; $e^{\rm BA}_{\rm det}$, the erroneous signal detection of Alice; $\rho^{\rm BE}$, the joint state after Eve's attack in the forward quantum channel; $Q_{\mu}^{\rm BAE}$, the overall signal gain of Eve; $\rho^{\rm BAE}$, the joint state after Alice's information encoding and Eve's attacks in the two quantum channels; $E_{\mu}^{\rm BAB}$ is QBER; $Q_{\mu}^{\rm BAB}$, the overall signal gain of Bob; $e^{\rm BAB}_{\rm det}$, the erroneous signal detection of Bob.}
\label{Fig:4}
\end{figure}

The attack strategies of Eve is shown in Fig. \ref{Fig:4}. Eve has the ability to discern the number of photons in every pulse, then the specific attack strategies performed by Eve would be divided into two types. On the one hand, if the pulse in the forward quantum channel contains only one photon $(n=1)$, Eve performs the collective attack on this photon \cite{lu2011unconditional,qi2019implementation}. To be more specific, Eve prepares ancilla states each of which interacts individually with the photons sent from Bob-to-Alice, and these ancilla states are stored in the quantum memory until the photons are returned from Alice after secret information has been encoded. Eve would perform the optimal measurement by combining her ancilla states and the encoded states in order to obtain the secret information. According to Ref. \cite{qi2019implementation}, the maximum information that Eve can obtain from a single photon is $I(A:E)_{n=1}=h(2e^{\rm BA}_{1})$, in which we have assumed reasonably that Eve introduces equivalent error rate in the $X$ and $Z$ basis, and $e^{\rm BA}_{1}$ is the DBER originated from a single photon. On the other hand, if the pulse in the forward quantum channel with photon numbers are greater than 1 $(n>1)$, Eve can perform the PNS attack. 

Let's start with the case $n\geq3$. The four linearly independent states $(\{|0\rangle^{\otimes n}, |1\rangle^{\otimes n}, |+\rangle^{\otimes n}, |-\rangle^{\otimes n}\}, n\geq3)$ could be unambiguously discriminated \cite{feng2004unambiguous}, hence there is a powerful attack that Eve can get all secret information for the pulse that contains multi-photon components $(n\geq3)$ and it goes as follows. Eve captures this pulse sent from Bob, then a new photon in the right state that is based on her successfully unambiguously discrimination would be prepared and transmitted to Alice. If Eve fails to discriminate the multi-photon state, she blocks it. After the secret information encoding is finished by Alice, Eve captures the pulse again and she can deterministically decode the secret information based on the known initial state. Consequently, the pulses with multiple photons $(n\geq3)$ referred to as multi-photon states cannot provide secrecy capacity in the DL04 QSDC protocol.

Indeed, $I(A\!\!:\!\!E)_{n\geq3}\!=\!1$, and we need to derive the secrecy capacity that two-photon components can achieve under the PNS attack. In the PNS attack, Eve splits one of the photons from the pulse that contains two photons in the forward quantum channel and retains it. As for the other photon, she applies the collective attack, as detailed above in the case of $n=1$. What is unusual is that Eve can get two intercepted photons from each pulse, and these states will be combined with her ancillas for the optimal measurement. We assume that the initial state prepared by Bob is $\rho_{B}=\left(|00\rangle\langle00|+|11\rangle\langle11|+|++\rangle\langle++|+|--\rangle\langle--|\right)/4$. Eve's quantum operation in the PNS attack can be represented as 
\begin{eqnarray}
&U&|0\rangle_{B}|0\rangle_{B}|E\rangle\!\!=\!\!|0\rangle_{B}|0\rangle_{B}|E_{0000}\rangle\nonumber+|0\rangle_{B}|1\rangle_{B}|E_{0001}\rangle\!\!=\!\!|\varphi_{1}\rangle,\nonumber\\
&U&|1\rangle_{B}|1\rangle_{B}|E\rangle\!\!=\!\!|1\rangle_{B}|0\rangle_{B}|E_{1110}\rangle\nonumber+|1\rangle_{B}|1\rangle_{B}|E_{1111}\rangle\!\!=\!\!|\varphi_{2}\rangle,\nonumber\\
&U&|+\rangle_{B}|+\rangle_{B}|E\rangle\!\!=\!\!|\varphi_{3}\rangle,\nonumber\\
&U&|-\rangle_{B}|-\rangle_{B}|E\rangle\!\!=\!\!|\varphi_{4}\rangle,
\end{eqnarray}
where $U$ is an unitary operation performed on two particles, i.e., one photon of $\rho_{B}$ together with $|E\rangle$ and $|E\rangle$ $(|E\rangle_{0000}, |E\rangle_{0001}, |E\rangle_{1110}$, and $|E\rangle_{1111})$ is the ancilla state before (after) attack. The effect of Alice's encoding unitary operation $Y$ (single-particle operation) on the photons can be written as 
\begin{eqnarray}
\label{Eq:collective attack2}
YU|0\rangle_{B}|0\rangle_{B}|E\rangle&=&-|0\rangle_{B}|1\rangle_{B}|E_{0000}\rangle\nonumber+|0\rangle_{B}|0\rangle_{B}|E_{0001}\rangle\nonumber\\
&=&|\varphi_{5}\rangle,\nonumber\\
YU|1\rangle_{B}|1\rangle_{B}|E\rangle&=&-|1\rangle_{B}|1\rangle_{B}|E_{1110}\rangle\nonumber+|1\rangle_{B}|0\rangle_{B}|E_{1111}\rangle\nonumber\\
&=&|\varphi_{6}\rangle,\nonumber\\
YU|+\rangle_{B}|+\rangle_{B}|E\rangle&=&|\varphi_{7}\rangle,\nonumber\\
YU|-\rangle_{B}|-\rangle_{B}|E\rangle&=&|\varphi_{8}\rangle,
\end{eqnarray}

Hence, after Eve's attack, the joint state of two photons and Eve's ancilla in the forward quantum channel is $\rho_{\rm BE}=U\left(\rho_{B}\otimes|E\rangle \langle E|\right)U^{\dagger}$. During the information encoding, if Alice performs unitary operation $I$ or $Y$ with the probability of $p_{0}$ and $p_{1}$ on the photons, respectively, the joint state would become $\rho^{0}_{\rm BE}=U\left(\rho_{B}\otimes|E\rangle \langle E|\right)U^{\dagger}$ or $\rho^{1}_{\rm BE}=YU\left(\rho_{B}\otimes|E\rangle \langle E|\right)U^{\dagger}Y^{\dagger}$ with respective probabilities. Thus, the joint state that Eve can access in the backward quantum channel is
\begin{eqnarray}
\label{Eq:jointstate}
\rho_{\rm BEA}&=&p_{0}\cdot\rho^{0}_{\rm BE}+p_{1}\cdot \rho^{1}_{\rm BE}\nonumber\\
&=&\frac{1}{4}(p_{0}|\varphi_{1}\rangle\langle\varphi_{1}|+p_{0}|\varphi_{2}\rangle\langle\varphi_{2}|+p_{0}|\varphi_{3}\rangle\langle\varphi_{3}|\nonumber\\
&+&p_{0}|\varphi_{4}\rangle\langle\varphi_{4}|+p_{1}|\varphi_{5}\rangle\langle\varphi_{5}|+p_{1}|\varphi_{6}\rangle\langle\varphi_{6}|\nonumber\\
&+&p_{1}|\varphi_{7}\rangle\langle\varphi_{7}|+p_{1}|\varphi_{8}\rangle\langle\varphi_{8}|),\nonumber\\
\end{eqnarray}
where $p_{0}+p_{1}=1$.

The maximum information that Eve can steal $I(A\!\!:\!\!E)$ is given by the Holevo bound $\chi$ \cite{holevo1973bounds,wu2019security}, that is,
\begin{eqnarray}
\label{Eq:Holevo}
I(A\!\!:\!\!E)&\leq&\chi \nonumber\\ &=&\max \limits_{\{U\}}\left\{S(\rho_{\rm BEA})-p_{0}\cdot S(\rho^{0}_{\rm BE})-p_{1}\cdot S(\rho^{1}_{\rm BE})\right\},\nonumber\\
\end{eqnarray}
where $S(\rho)=-{\rm Tr}(\rho {\rm log}_{2}\rho)$ represents the von Neumann entropy. On the one hand, since the density operator $\rho^{0}_{\rm BE}$ and $\rho^{1}_{\rm BE}$ are only different in unitary operation from $\rho_{B}\otimes|E\rangle \langle E|$, we can obtain that $S(\rho^{0}_{\rm BE})=S(\rho^{1}_{\rm BE})=S(\rho_{B}\otimes|E\rangle \langle E|)=3/2$. On the other hand, we must obtain the eigenvalues of the joint state $\rho_{\rm BEA}$ in order to calculate the von Neumann entropy $S(\rho_{\rm BEA})$. We can simplify the process of calculating eigenvalues by using the Gram matrix representation, which is proved to have the same eigenvalues with its corresponding density operator \cite{jozsa2000distinguishability}. For the joint state $\rho_{\rm BEA}$, its Gram matrix is given by
\begin{eqnarray}
\label{Eq:Gram}
G\!\!=\!\!\frac{1}{4}\!\!\begin{bmatrix}
p_{0}\langle\varphi_{1}|\varphi_{1}\rangle & \!\!\!p_{0}\langle\varphi_{1}|\varphi_{2}\rangle  &\!\!\!\! \cdots   &\!\!\!\! \sqrt{p_{0}p_{1}}\langle\varphi_{1}|\varphi_{8}\rangle   \\

p_{0}\langle\varphi_{2}|\varphi_{1}\rangle & \!\!\!p_{0}\langle\varphi_{2}|\varphi_{2}\rangle  &\!\!\!\! \cdots   &\!\!\!\! \sqrt{p_{0}p_{1}}\langle\varphi_{2}|\varphi_{8}\rangle  \\

\vdots & \vdots  &\!\!\!\!\!\! \ddots   &\!\!\!\!\! \vdots  \\

\sqrt{p_{0}p_{1}}\langle\varphi_{8}|\varphi_{1}\rangle & \!\!\!\sqrt{p_{0}p_{1}}\langle\varphi_{8}|\varphi_{2}\rangle  &\!\!\!\! \cdots  &\!\!\!\! p_{1}\langle\varphi_{8}|\varphi_{8}\rangle  
\end{bmatrix}
.\nonumber\\
\end{eqnarray}
Note that the above analysis applies to the most general PNS attack. To illustrate the use of the above result, we assume that Eve's attack operator $U$ is symmetric, which further means that her attack could be modeled as a depolarizing channel \cite{krawec2017quantum}. The depolarizing channel is a typical model invoked in the unconditional security proofs of some QKD protocols, as detailed in \cite{christandl2004generic,scarani2009security,henao2015practical}. Hence, in addition to the conditions of orthonormality, $\langle E_{0000}|E_{0000}\rangle+\langle E_{0001}|E_{0001}\rangle=1$ and $\langle E_{1110}|E_{1110}\rangle+\langle E_{1111}|E_{1111}\rangle=1$, there are some equations of the depolarizing channel to calculate the specific values of Gram matrix's elements, which are given as follows \cite{henao2015practical,qi2019implementation}
\begin{eqnarray}
\label{Eq:dc}
\langle E_{0000}|E_{1110}\rangle=\langle E_{0001}|E_{1111}\rangle=0,\nonumber\\ 
\langle E_{0000}|E_{0001}\rangle=\langle E_{1110}|E_{1111}\rangle=0,\nonumber\\
\langle E_{0001}|E_{1110}\rangle=0,~~~~~~~~~~~~~~~~~~~~~~~~~~~~\nonumber\\ 
\langle E_{0000}|E_{1111}\rangle=1-2e^{\rm BA}_{2},~~~~~~~~~~~~~~~
\end{eqnarray}
where $e^{\rm BA}_{2}$ is the DBER caused by two photons from Bob-to-Alice. Furthermore, we assume that $p_{0}=p_{1}=1/2$ \cite{qi2019implementation}. After cumbersome calculations, we can get that the eigenvalues of $\rho_{\rm BEA}$ are $\lambda^{\rm BEA}_{1,2}=0$, $\lambda^{\rm BEA}_{3,4}=1/4$, $\lambda^{\rm BEA}_{5,6}=(1-2e^{\rm BA}_{2})/4$, and $\lambda^{\rm BEA}_{7,8}=2e^{\rm BA}_{2}/4$. Therefore, $S(\rho_{\rm BEA})=-{\rm Tr}(\rho_{\rm BEA} {\rm log}_{2}\rho_{\rm BEA})=-\sum_{\textit{i}}\lambda^{\rm BEA}_{\textit{i}}{\rm log}_{2}(\lambda^{\rm BEA}_{\textit{i}})=2+\textit{h}(2\textit{e}_{2})/2$, where $h(x)=-x{\rm log}_{2}(x)-(1-x){\rm log}_{2}(1-x)$ is the binary Shannon entropy. According to Eq. (\ref{Eq:Holevo}), the maximum information that Eve can steal via the pulse containing two photons is 
\begin{equation}
\label{Eq:IAE2}
I(A:E)_{n=2}=\frac{1}{2}h(2e^{\rm BA}_{2})+\frac{1}{2}.
\end{equation}
One important conclusion we can draw from Eq. (\ref{Eq:IAE2}) is that the DL04 QSDC protocol \cite{deng2004secure} has the ability to defend against the PNS attack in the case of two photons, since $I(A:E)_{n=2}$ could be below 1. The basic physics is that no basis announcement is required in QSDC for information decoding, while basis comparison is necessary for establishing the common secret keys in the BB84 QKD \cite{bennett1984quantum}. 

\subsection{System model}
\label{subsec:System model}
In order to analyze the practical QSDC experiment system, let us calculate $I(A:B)$ and $I(A:E)$ under the frame of Eve performing the general collective attack on single photon and the PNS attack on multi photons, considering the device and channel losses. Assuming that $\alpha^{\rm BA}$ and $\alpha^{\rm BAB}$ are the channel attenuation of different paths $\rm BA$ and $\rm BAB$, respectively. As can be seen in Fig. \ref{Fig:4}, Eve performs her eavesdropping after Alice finishing information encoding, which indicates $\alpha^{\rm BAB}=2\alpha^{\rm BA}$. Thus, we have the channel transmissions as follows:
\begin{eqnarray}
t^{\rm BA}=10^{-\left(\frac{\alpha^{\rm BA}}{10}\right)},\nonumber\\
t^{\rm BAB}=10^{-\left(\frac{\alpha^{\rm BAB}}{10}\right)},
\end{eqnarray}
then the concomitant overall transmissions are given by
\begin{eqnarray}
\eta^{\rm BA}=t^{\rm BA}\eta_{\rm opt}^{\rm BA}\eta^{A}_{D},\nonumber\\
\eta^{\rm BAB}=t^{\rm BAB}\eta_{\rm opt}^{\rm BAB}\eta^{B}_{D},
\end{eqnarray}
where $\eta_{\rm opt}^{\rm BA}$ and $\eta_{\rm opt}^{\rm BAB}$ are the specific devices' intrinsic optical losses, while $\eta^{A}_{D}$ and $\eta^{B}_{D}$ are the detection efficiency of Alice and Bob, respectively. The transmittances of $n$-photon state through different paths are $\eta^{\rm BA}_{n}=1-(1-\eta^{\rm BA})^{n}$ and $\eta^{\rm BAB}_{n}=1-(1-\eta^{\rm BAB})^{n}$. With $Y^{A}_{0}$ and $Y^{B}_{0}$ as background detection events of different parties, the yields become $Y^{A}_{n}=Y^{A}_{0}+\eta^{\rm BA}_{n}-\eta_{n}^{\rm BA} Y^{A}_{0}\approx Y^{A}_{0}+\eta^{\rm BA}_{n}$ and $Y^{B}_{n}\approx Y^{B}_{0}+\eta^{\rm BAB}_{n}$, the overall signal gains and the error rates are given by \cite{ma2005practical}:
\begin{eqnarray}
\label{Eq:allQ}
Q^{\rm BA}_{\mu}&=&\sum_{n=0}^{\infty}Q^{\rm BA}_{\mu,n}\!=\!\sum_{n=0}^{\infty}p(n,\mu)Y^{A}_{n}\nonumber\\
&=&Y^{A}_{0}+1-e^{-\eta^{\rm BA}\mu},\nonumber\\
Q^{\rm BAE}_{\mu}&=&\sum_{n=0}^{\infty}\!\!Q^{\rm BAE}_{\mu,n}\nonumber\\
&\leq&\sum_{n=0}^{\infty}\left [Q_{\mu,n}^{\rm BA}\!-\!p(n,\mu)Y_{0}^{A}  \right ]\!\max\left\{1,\frac{\gamma^{E}}{\gamma^{A}}\right\},\nonumber\\
Q^{\rm BAB}_{\mu}&=&\sum_{n=0}^{\infty}Q^{\rm BAB}_{\mu,n}\!=\!\sum_{n=0}^{\infty}p(n,\mu)Y^{B}_{n}\nonumber\\
&=&Y^{B}_{0}+1-e^{-\eta^{\rm BAB}\mu},\nonumber\\
\end{eqnarray}
and
\begin{eqnarray}
E^{\rm BA}_{\mu}=\frac{e_{0}Y^{A}_{0}+e^{\rm BA}_{\rm det}(1-e^{-\eta^{\rm BA}\mu})}{Q^{\rm BA}_{\mu}},\nonumber\\
E^{\rm BAB}_{\mu}=\frac{e_{0}Y^{B}_{0}+e^{\rm BAB}_{\rm det}(1-e^{-\eta^{\rm BAB}\mu})}{Q^{\rm BAB}_{\mu}},
\end{eqnarray}
where $e_{0}=1/2$ is the error rate of background, $Q^{\rm BA}_{\mu,n}$ ($Q_{\mu,n}^{\rm BAE}$ and $Q_{\mu,n}^{\rm BAB}$) is the $n$-photon signal gain at Alice (Eve and Bob), and $e^{\rm BA}_{\rm det}$ as well as $e^{\rm BAB}_{\rm det}$ are intrinsic detector error rates which can be calculated by the visibilities $V$ of the detection system: $e^{\rm BA}_{\rm det}=(1-V^{\rm BA})/2$ and $e^{\rm BAB}_{\rm det}=(1-V^{\rm BAB})/2$ \cite{ma2008quantum}. The derivation of $Q^{\rm BAE}_{\mu,n}$ is given in the Appendix.

According to the theory of binary symmertric channel and binary erasure channel \cite{mackay2003information}, the mutual information between Alice and Bob can be calculated as
\begin{eqnarray}
I(A:B)=Q_{\mu}^{\rm  BAB}\left[ 1-h(E^{\rm  BAB}_{\mu})\right],
\end{eqnarray}
where $Q_{\mu}^{\rm  BAB}$ is the overall signal gain of Bob after a round trip $\rm BAB$, and $E^{\rm BAB}_{\mu}$ is the QBER. The secret information that Eve can obtain from single photon by using the collective attack is \cite{lu2011unconditional,qi2019implementation} 
\begin{eqnarray}
I(A:E)_{n=1}=Q_{\mu,n=1}^{\rm BAE}h(2e^{\rm BA}_{1}),
\end{eqnarray}
where $e^{\rm BA}_{1}$ is the DBER caused by the single photon. Given the above, the lower bound of secrecy capacity is
\begin{eqnarray}
\label{Eq:systemC}
C_{s}&=&Q_{\mu}^{\rm BAB}\left[ 1-h(E^{\rm BAB}_{\mu})\right]-Q_{\mu,n=1}^{\rm BAE}h(2e^{\rm BA}_{1})\nonumber\\
&-&Q_{\mu,n=2}^{\rm BAE}\left[\frac{1}{2}h(2e^{\rm BA}_{2})+\frac{1}{2}\right]-Q_{\mu,n\geq3}^{\rm BAE}\cdot1.
\end{eqnarray}
Obviously, now we need to discuss how to evaluate the DBERs in the Eq. (\ref{Eq:systemC}) caused by single-photon ($e^{\rm BA}_{1}$) states and two-photon ($e^{\rm BA}_{2}$) states.

\subsection{GLLP theory}
\label{subsec:GLLP theory}
There is a pessimistic assumption in the GLLP theory \cite{gottesman2004security}: all multi-photon signals could be detected by Alice and all errors originate from the single photon. Hence, the upper bound of $e^{\rm BA}_{1}$ is evaluated by
\begin{eqnarray}
e^{\rm BA}_{1}=\frac{E^{\rm  BA}_{\mu}}{1-\frac{p(n\geq2,\mu)}{Q_{\mu}^{\rm  BA}}},
\end{eqnarray}
where $E^{\rm BA}_{\mu}$ is the DBER and $Q_{\mu}^{\rm BA}$ is the overall signal gain at Alice's terminal after the $\rm BA$ path. However, the GLLP theory cannot give us a real value of $e^{\rm BA}_{2}$, in other words, $e^{\rm BA}_{2}=0$ with its assumption. In this case, $I(A:E)_{n=2}=Q_{\mu,n=2}^{\rm BAE}\cdot(1/2)$ according to Eq. (\ref{Eq:IAE2}) and Eq. (\ref{Eq:systemC}), which means Eve can obtain a part of secret information from two-photon state by the zero-DBER eavesdropping. Actually, it is a special case of our PNS attack. Eve intercepts one photon in the forward quantum channel but does nothing for the other and forwards it directly (no error rate here, $e^{\rm BA}_{2}=0$). After Alice finishes secret information encoding, Eve intercepts the encoded photon and combines the intercepted two photons to read secret information. Note that the PNS attack needs to be combined with the unambiguous state discrimination (USD) attack \cite{zhang2002set}, namely, Eve obtains information by discriminating the states before and after Alice's encoding operation, since there is no basis reconciliation in the DL04 QSDC protocol \cite{deng2004secure}. The upper bound on the maximum probability to discriminate two mixed states is 1/2 \cite{lin2009eavesdropping}, which matches the above-mentioned result $I(A:E)_{n=2}=Q_{\mu,n=2}^{\rm BAE}\cdot(1/2)$ we have obtained under the PNS attack, in which the secret information Eve may steal from two-photon state is 1/2 without considering her reception rate $Q_{\mu,n=2}^{\rm BAE}$. Based on the assumption of GLLP, the value of $Q_{\mu,n=1}^{\rm BAE}\!\!=\!\!Q_{\mu}^{\rm BA}-p(n\geq2,\mu)-p(0,\mu)Y_{0}^{A}-p(1,\mu)Y_{0}^{A}$, $Q_{\mu,n=2}^{\rm BAE}=p(2,\mu)-p(2,\mu)Y_{0}^{A}$, and $Q_{\mu,n\geq3}^{\rm BAE}\!\!=\!\!p(n\geq3,\mu)-p(n\geq3,\mu)Y_{0}^{A}$ in GLLP can be estimated by combining the Eq. (\ref{Eq:QBAEn}) and the constraint of the first formula of Eq. (\ref{Eq:allQ}) for maximizing $I(A:E)$.

\subsection{Decoy state method}
\label{subsec:Decoy state method}
One way to beat the PNS attack in QKD is by utilizing decoy state method \cite{hwang2003quantum,wang2005beating,lo2005decoy}. This method also can be integrated into the DL04 QSDC \cite{deng2004secure}, and we consider the decoy state here only for detecting the PNS attack, leaving the problem of whether it can be used to transmit secret information for future work. More importantly, the decoy state can provide a better estimation of the DBER. Bob randomly uses the signal source or the decoy source to prepare the initial states and sends them to Alice. Once these states are received by Alice, she randomly chooses some of them to publicly discuss with Bob for eavesdropping detection that is the same as Step (2) in Section \ref{subsec:Protocol}. Bob announces where the decoy states are and then the transmission properties would be tested by Alice. It is impossible for Eve to discriminate which ones are the decoy states, in this way, if Eve still performs the PNS attack in the forward quantum channel, the counting rate of the system in path of $\rm BA$ will be inevitably disturbed. If Alice and Bob confirm that the forward quantum channel has not been tapped, Alice will use the remaining signal states for information encoding. 

Much of the decoy-state research in the Scarani-Acin-Ribordy-Gisin 2004 (SARG 04) QKD protocol \cite{scarani2004quantum,fung2006performance,zhang2007limitation,jing2006nonorthogonal} has shown how the decoy state method can be used to estimate the error rate caused by two photons. Inspired by these previous works, we use four decoy states: one vacuum state and three weak decoy states ($\nu_{1}$, $\nu_{2}$, and $\nu_{3}$) to estimate our $e^{\rm BA}_{2}$, so that the background rate can be estimated by the vacuum state, i.e., $Y^{A}_{0}=Q^{\rm BA}_{\rm vac}$ and $e_{0}=E^{\rm BA}_{\rm vac}=1/2$. The upper bound of single-photon DBER and two-photon DBER are, respectively, given by \cite{jing2006nonorthogonal}
\begin{equation}
e^{BA,U}_{1}=\frac{E^{\rm BA}_{\nu_{3}}Q^{\rm BA}_{\nu_{3}}e^{\nu_{3}}-e_{0}Y^{A}_{0}}{Y^{A,L}_{1}\nu_{3}}
\end{equation}
and
\begin{equation}
e^{BA,U}_{2}\!=\!\frac{2\left(E_{\nu_{2}}^{\rm BA}Q_{\nu_{2}}^{\rm BA}e^{\nu_{2}}\!-\!\frac{\nu_{2}}{\nu_{3}}E_{\nu_{3}}^{\rm BA}Q_{\nu_{3}}^{\rm BA}e^{\nu_{3}}\!+\!\frac{\nu_{2}-\nu_{3}}{\nu_{3}}e_{0}Y_{0}^{A}\right)}{Y^{A,L}_{2}\nu_{2}\left (\nu_{2}-\nu_{3}\right )},
\end{equation}
where
\begin{equation}
Y^{A,L}_{1}\!=\!\frac{\mu^2\left(Q^{\rm BA}_{\nu_{2}}e^{\nu_{2}}\!-\!Q^{\rm BA}_{\nu_{3}}e^{\nu_{3}}\right)\!-\!\left(\nu_{2}^2-\nu_{3}^2\right)\left(Q^{\rm BA}_{\mu}e^{\mu}\!-\!Y^{A}_{0}\right)}{\mu\left(\nu_{2}\!-\!\nu_{3}\right)\left(\mu\!-\!\nu_{2}\!-\!\nu_{3}\right)}
\end{equation}
and
\begin{equation}
Y^{A,L}_{2}\!=\!\frac{2\mu\left(Q^{\rm BA}_{\nu_{1}}e^{\nu_{1}}\!\!-\!Q^{\rm BA}_{\nu_{2}}e^{\nu_{2}}\right)\!\!-\!\!2\left(\nu_{1}\!\!-\!\!\nu_{2}\right)\left(Q^{\rm BA}_{\mu}e^{\mu}\!\!-\!Y^{A}_{0}\right)}{\mu\left(\nu_{1}\!\!-\!\nu_{2}\right)\left(\nu_{1}\!\!+\!\nu_{2}\!\!-\!\!\mu\right)}. 
\end{equation}
Furthermore, the above mean photon numbers $\mu$, $\nu_{1}$, $\nu_{2}$ and $\nu_{3}$ meet the following conditions
\begin{eqnarray}
&&0<\nu_{3}<\nu_{2}\leq \frac{2}{3} \mu<\nu_{1}\leq\frac{3}{4}\mu,\nonumber\\
&&\nu_{1}+\nu_{2}>\mu,\nonumber\\
&&\nu_{2}+\nu_{3}<\mu,\nonumber\\
&&\nu_{1}-\nu_{2}-\frac{\nu_{1}^{3}-\nu_{2}^{3}}{\mu^2}=0.
\end{eqnarray}
Results with explicit examples obtained from Eq. (\ref{Eq:systemC}) are given in Fig. \ref{Fig:6}.

\subsection{Performance analysis}
\label{subsec:Performance analysis}
The devices' intrinsic optical losses are measured from our experimental setup. There is an altogether loss of 4.3 dB from PBS and PM. The attenuation of the short-arm optical link of the Mach-Zehnder interferometer is 2.3 dB. Suppose Eve's detection efficiency is $\eta^{E}_{D}=100\%$ and without background detection events, while Alice and Bob utilize the superconducting single-photon detector with detection efficiency $\eta^{A}_{D}=\eta^{B}_{D}=70\%$ and background detection events $Y^{A}_{0}=Y^{B}_{0}=8\times10^{-8}$. $\gamma^{A}=(1-k)\cdot10^{-2.3/10}\cdot70\%$ and $\gamma^{E}=k\cdot10^{-4.3/10}\cdot k\cdot100\%$, where $k$ originates from a (1-k):k BS. Then, the overall device intrinsic optical loss of Alice and Bob are given by $\eta_{\rm opt}^{\rm BA}=(1-k)\cdot10^{-2.3/10}$ and $\eta_{\rm opt}^{\rm BAB}=k^2\cdot10^{-6.6/10}$, respectively. The intrinsic detector error rates $e^{\rm BA}_{\rm det}=1.31\%$ and $e^{\rm BAB}_{\rm det}=0.26\%$ are deduced from system visibilities. Furthermore, the value of k is fixed by $\gamma^{A}=\gamma^{E}$. We then performed a numerical simulation to estimate the secrecy capacity under Eve's attacks with this setup in terms of maximum optical link attention.

\begin{figure}[htbp]%
\centering\includegraphics[width=8.5cm]{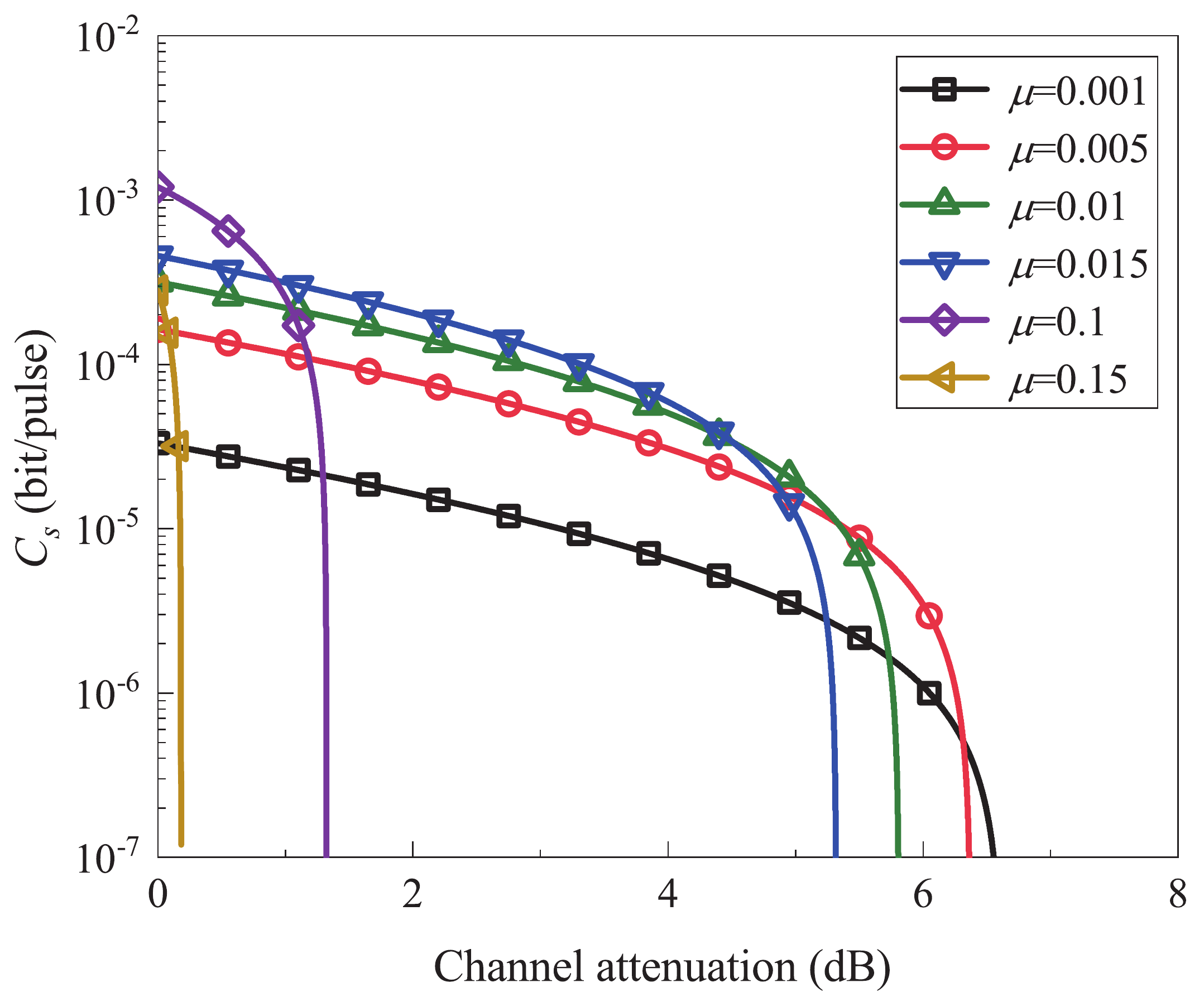}
\caption{Secrecy capacities versus the attenuation given the collective attack as well as the PNS and USD attack under the framework of GLLP analysis. The curves labeled by different markers represent the data with different mean photon numbers.}
\label{Fig:5}
\end{figure}

Figure \ref{Fig:5} shows the secrecy capacity of the free-space QSDC system with different mean photon numbers given by the GLLP theory. There is a trade-off between the secrecy capacity and the maximum tolerable attenuation. The maximum tolerable attenuation would be very small with the large mean photon numbers due to the high multi-photon probability in pulse, and it is susceptible to the PNS attack. However, it is infeasible to improve the maximum tolerable attenuation by reducing the mean photon numbers drastically on account of the decrease in the secrecy capacity. Hence, we choose the mean photon number $\mu=0.01$ as the near-optimal value to highlight performance, as this is its preferable performance both in the secrecy capacity and in the maximum tolerable attenuation. Consequently, as shown in Fig. \ref{Fig:5}, the channel attenuation of secure communication against the collective attack as well as the PNS and USD attack for the QSDC system with realistic devices is less than 5.8 dB.

\begin{figure}[htbp]%
\centering\includegraphics[width=8.5cm]{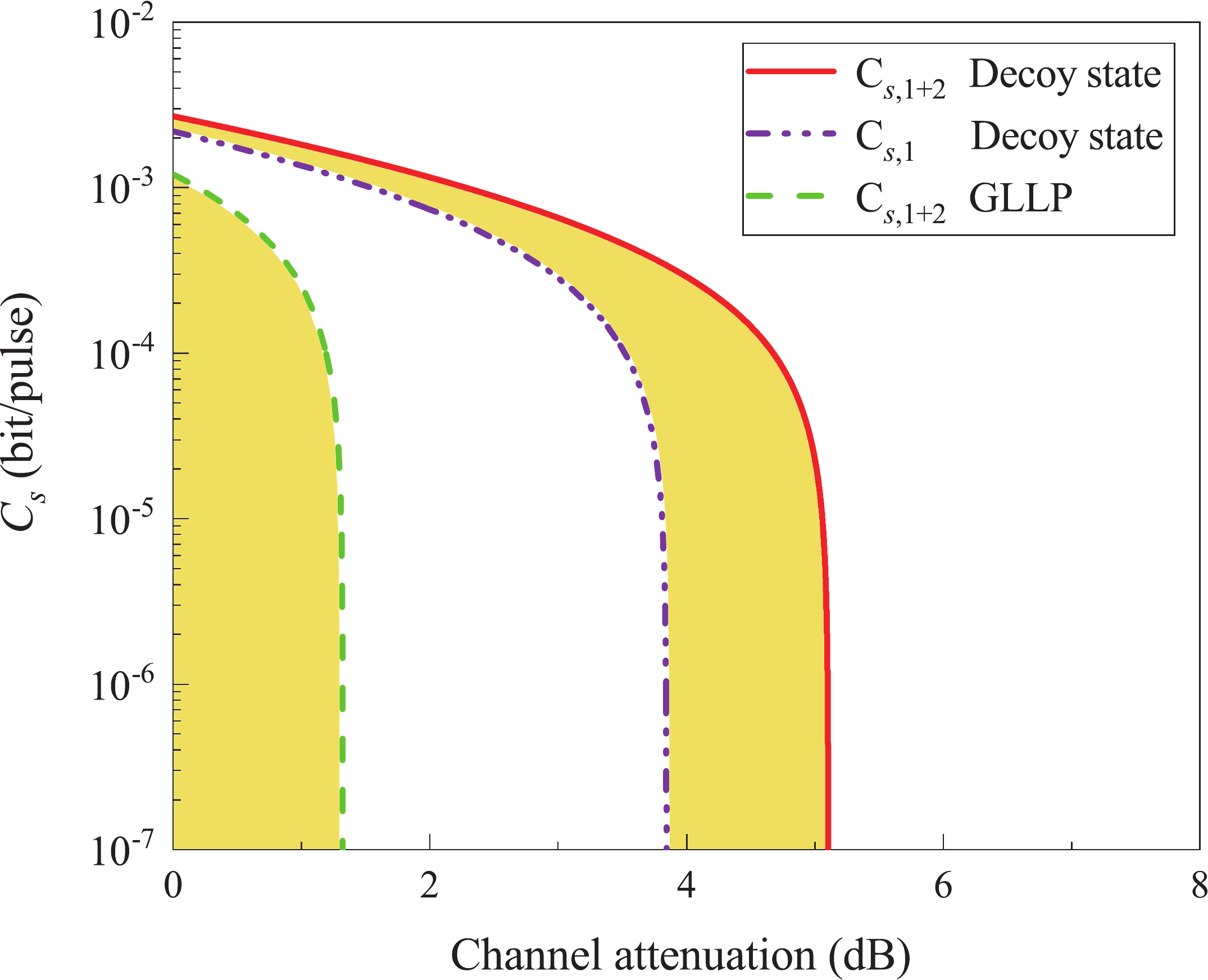}
\caption{Comparision of the secrecy capacities calculated by GLLP theory and decoy state method. Simulation in the decoy state method using $\mu=0.1$, $\nu_{1}=0.07$, $\nu_{2}=0.0445$, and $\nu_{3}=0.03$ and in the GLLP using $\mu=0.1$. In the secrecy capacity $C_{s,1+2}$, we have considered the contribution both from single-photon states and two-photon states, while $C_{s,1}$ has not considered the contribution from two-photon states. The two yellow areas represent the contribution of two-photon states to the secrecy capacity.}
\label{Fig:6}
\end{figure}

By contrast, as shown in Fig. \ref{Fig:6}, the secrecy capacity and the maximum tolerable attenuation can be greatly increased by using decoy state method. To be more specific, the maximum tolerable attenuation of decoy state method is 3.9 times than that of GLLP. The results show that the decoy state can accurately estimate the DBER caused by single- and two-photon state in which it plays a positive role in improving communication performance, rather than the GLLP theory that gives a poor estimation. As seen in Fig. \ref{Fig:6}, the contribution of a two-photon state to the secrecy capacity cannot be completely disregarded, especially when the system is operated with a comparatively higher mean photon number. For GLLP, there is even no secrecy capacity at $\mu=0.1$ if the contribution of two-photon components has not been considered.

In clear weather conditions, the typical atmosphere attenuation is 0.5\textasciitilde2 dB/km \cite{kim2001availability,carrasco2014correction}, it is feasible to exchange secret information by free-space QSDC based on phase-encoding for two users over more than 1 km without using decoy state, which is a typical distance between two terminals in a secure area. If the decoy state method is applied, this secure communication distance could be further improved. One typical usage scenario would be applied in indoor environments for wireless communication, known as the quantum Li-Fi system \cite{elmabrok2018wireless}.

\section{Conclusions}
\label{sec:Conclusions}
We have constructed a free-space QSDC system based on phase encoding. The asymmetric Mach-Zehnder interferometers serve as transmitter and receiver with convincing fringe visibilities. The system can be operated to transmit text, picture, and audio, with a low average QBER of 0.49\%$\pm$0.27\%. This indicates the feasibility of phase-encoding based QSDC over a free-space channel. The security analysis of free-space QSDC has been given under the general collective on single-photon state and the PNS attack on multi-photon state, making a beneficial step to calculate the secrecy capacity of QSDC system using a practical light source. Furthermore, the PNS attack is a general strategy that applicable to explain the previous PNS plus USD attack \cite{lin2009eavesdropping}. Our results show that the DL04 QSDC protocol is robust against the PNS attack in the depolarizing channel, and the secrecy capacity is increased significantly after considering the security of two-photon components, especially under the framework of decoy state. As for future investigation, the effects of background light noise need to be considered in the free-space QSDC system. Decreasing the intrinsic loss of optical setups, and optimizing the decoy state method will be beneficial for long-distance transmission of QSDC over a free-space channel. It is worth mentioning that the phase drift of photon must be carefully handled by the free-space QSDC system with phase-encoding. Hence, the maximum communication distance of free-space QSDC with phase-encoding needs to be further investigated.

\appendix* 
\section{A}
\label{sec:Appendix}
We can estimate $Q_{\mu,n}^{\rm BAE}$ from the value of $Q_{\mu,n}^{\rm BA}$, since they are related to the number of photons received by Alice. For $n$ photons emitted by Bob, Alice actually receives $m$ photons at her port BS after the forward quantum channel. The photon number distribution is $f_{n}(m,\mu)$, which is no longer a Poissonian distribution under the PNS attack. The yields of Alice and Eve for these photons, are, respectively, given by
\begin{eqnarray}
\label{eq:ApendQBA}
Y_{n}^{A}\!-\!Y_{0}^{A}\!\!&=&\!\!\sum_{m=0}^{\infty}\!\!f_{n}(m,\mu)\!\left[1\!-\!(1\!-\!\gamma^{A})^m\!-\!(1\!-\!(1\!-\!\gamma^{A})^m)Y_{0}^{A}\right]\nonumber\\
\!\!&\approx&\!\!\sum_{m=0}^{\infty}\!f_{n}(m,\mu)\!\left[1\!-\!(1\!-\!\gamma^{A})^m\right],
\end{eqnarray}
and
\begin{eqnarray}
\label{eq:ApendQBE}
Y^{E}_{n}\!\!&=&\!\!\sum_{m=0}^{\infty}\!\!f_{n}(m,\mu)\!\left[1\!-\!(1\!-\!\gamma^{E})^m\!\!-\!(1\!-\!(1\!-\!\gamma^{E})^m)Y_{0}^{E}\right]\!\!+\!\!Y_{0}^{E}\nonumber\\
\!\!&\approx&\!\!\sum_{m=0}^{\infty}\!f_{n}(m,\mu)\!\left[1\!-\!(1\!-\!\gamma^{E})^m\right],
\end{eqnarray}
where $\gamma^{A}$ is the overall transmission for photons received and then measured by Alice, $\gamma^{E}$ is the overall transmission of Eve after Alice encodes her receiving photons, and $Y_{0}^{E}=0$. Combining Eq. (\ref{eq:ApendQBA}) and Eq. (\ref{eq:ApendQBE}), the yields of Eve $Y^{E}_{n}$ becomes
\begin{eqnarray}
\label{eq:AppendQBE_F}
Y^{E}_{n}&=&(Y_{n}^{A}\!-\!Y_{0}^{A})\frac{\sum\limits_{m=0}^{\infty}\!f_{n}(m,\mu)\!\left[1\!-\!(1\!-\!\gamma^{E})^m\right]}{\sum\limits_{m=0}^{\infty}\!f_{n}(m,\mu)\!\left[1\!-\!(1\!-\!\gamma^{A})^m\right]}\nonumber\\
&\leq&(Y_{n}^{A}\!-\!Y_{0}^{A})\max\left\{1,\frac{\gamma^{E}}{\gamma^{A}}\right\},
\end{eqnarray}
where we have utilized the following mathematical property
\begin{eqnarray}
\begin{cases}
 \frac{\sum\limits_{m=0}^{\infty}\!f_{n}(m,\mu)\!\left[1\!-\!(1\!-\!\gamma^{E})^m\right]}{\sum\limits_{m=0}^{\infty}\!f_{n}(m,\mu)\!\left[1\!-\!(1\!-\!\gamma^{A})^m\right]}\leq1 & \text{ if } \gamma^{A}\geq \gamma^{E}\\ 
 \frac{\sum\limits_{m=0}^{\infty}\!f_{n}(m,\mu)\!\left[1\!-\!(1\!-\!\gamma^{E})^m\right]}{\sum\limits_{m=0}^{\infty}\!f_{n}(m,\mu)\!\left[1\!-\!(1\!-\!\gamma^{A})^m\right]}\leq \frac{\gamma^{E}}{\gamma^{A}}& \text{ if }  \gamma^{A}<\gamma^{E}
\end{cases}.
\end{eqnarray}
The gain of the $n$-photon state of Alice and Eve are $Q_{\mu,n}^{\rm BA}=p(n,\mu)Y_{n}^{A}$ and $Q^{\rm BAE}_{\mu,n}=p(n,\mu)Y_{n}^{E}$, respectively. Hence, we have
\begin{eqnarray}
\label{Eq:QBAEn}
Q^{\rm BAE}_{\mu,n}\!=\!p(n,\mu)Y_{n}^{E}\leq \left [Q_{\mu,n}^{\rm BA}\!-\!p(n,\mu)Y_{0}^{A}    \right]\max\left\{1,\frac{\gamma^{E}}{\gamma^{A}}\right\}.\nonumber\\
\end{eqnarray}

\section*{Funding}
Government of Guangdong province (2018B030325002); National Natural Science Foundation of China (11974205); Ministry of Science and Technology of the People’s Republic of China (2017YFA0303700); Beijing Advanced Innovation Center for Future Chip (ICFC).

\bibliographystyle{apsrev4-1}
\bibliography{ref}

\begin{thebibliography}{63}%
\makeatletter
\providecommand \@ifxundefined [1]{%
 \@ifx{#1\undefined}
}%
\providecommand \@ifnum [1]{%
 \ifnum #1\expandafter \@firstoftwo
 \else \expandafter \@secondoftwo
 \fi
}%
\providecommand \@ifx [1]{%
 \ifx #1\expandafter \@firstoftwo
 \else \expandafter \@secondoftwo
 \fi
}%
\providecommand \natexlab [1]{#1}%
\providecommand \enquote  [1]{``#1''}%
\providecommand \bibnamefont  [1]{#1}%
\providecommand \bibfnamefont [1]{#1}%
\providecommand \citenamefont [1]{#1}%
\providecommand \href@noop [0]{\@secondoftwo}%
\providecommand \href [0]{\begingroup \@sanitize@url \@href}%
\providecommand \@href[1]{\@@startlink{#1}\@@href}%
\providecommand \@@href[1]{\endgroup#1\@@endlink}%
\providecommand \@sanitize@url [0]{\catcode `\\12\catcode `\$12\catcode
  `\&12\catcode `\#12\catcode `\^12\catcode `\_12\catcode `\%12\relax}%
\providecommand \@@startlink[1]{}%
\providecommand \@@endlink[0]{}%
\providecommand \url  [0]{\begingroup\@sanitize@url \@url }%
\providecommand \@url [1]{\endgroup\@href {#1}{\urlprefix }}%
\providecommand \urlprefix  [0]{URL }%
\providecommand \Eprint [0]{\href }%
\providecommand \doibase [0]{http://dx.doi.org/}%
\providecommand \selectlanguage [0]{\@gobble}%
\providecommand \bibinfo  [0]{\@secondoftwo}%
\providecommand \bibfield  [0]{\@secondoftwo}%
\providecommand \translation [1]{[#1]}%
\providecommand \BibitemOpen [0]{}%
\providecommand \bibitemStop [0]{}%
\providecommand \bibitemNoStop [0]{.\EOS\space}%
\providecommand \EOS [0]{\spacefactor3000\relax}%
\providecommand \BibitemShut  [1]{\csname bibitem#1\endcsname}%
\let\auto@bib@innerbib\@empty
\bibitem [{\citenamefont {Sun}\ \emph {et~al.}(2014)\citenamefont {Sun},
  \citenamefont {Liu}, \citenamefont {Lin}, \citenamefont {Zhang},
  \citenamefont {Lv}, \citenamefont {Huang}, \citenamefont {Huo}, \citenamefont
  {Yang}, \citenamefont {Jenkins}, \citenamefont {Zhao},\ and\ \citenamefont
  {Huang}}]{sun2014smart}%
  \BibitemOpen
  \bibfield  {author} {\bibinfo {author} {\bibfnamefont {H.}~\bibnamefont
  {Sun}}, \bibinfo {author} {\bibfnamefont {S.}~\bibnamefont {Liu}}, \bibinfo
  {author} {\bibfnamefont {W.}~\bibnamefont {Lin}}, \bibinfo {author}
  {\bibfnamefont {K.~Y.}\ \bibnamefont {Zhang}}, \bibinfo {author}
  {\bibfnamefont {W.}~\bibnamefont {Lv}}, \bibinfo {author} {\bibfnamefont
  {X.}~\bibnamefont {Huang}}, \bibinfo {author} {\bibfnamefont
  {F.}~\bibnamefont {Huo}}, \bibinfo {author} {\bibfnamefont {H.}~\bibnamefont
  {Yang}}, \bibinfo {author} {\bibfnamefont {G.}~\bibnamefont {Jenkins}},
  \bibinfo {author} {\bibfnamefont {Q.}~\bibnamefont {Zhao}}, \ and\ \bibinfo
  {author} {\bibfnamefont {W.}~\bibnamefont {Huang}},\ }\href@noop {}
  {\bibfield  {journal} {\bibinfo  {journal} {Nature communications}\ }\textbf
  {\bibinfo {volume} {5}},\ \bibinfo {pages} {3601} (\bibinfo {year}
  {2014})}\BibitemShut {NoStop}%
\bibitem [{\citenamefont {Cai}\ \emph {et~al.}(2017)\citenamefont {Cai},
  \citenamefont {Shi}, \citenamefont {Li}, \citenamefont {Gu}, \citenamefont
  {Ni}, \citenamefont {Cheng}, \citenamefont {Wang}, \citenamefont {Xiong},
  \citenamefont {Li}, \citenamefont {An},\ and\ \citenamefont
  {Huang}}]{cai2017visible}%
  \BibitemOpen
  \bibfield  {author} {\bibinfo {author} {\bibfnamefont {S.}~\bibnamefont
  {Cai}}, \bibinfo {author} {\bibfnamefont {H.}~\bibnamefont {Shi}}, \bibinfo
  {author} {\bibfnamefont {J.}~\bibnamefont {Li}}, \bibinfo {author}
  {\bibfnamefont {L.}~\bibnamefont {Gu}}, \bibinfo {author} {\bibfnamefont
  {Y.}~\bibnamefont {Ni}}, \bibinfo {author} {\bibfnamefont {Z.}~\bibnamefont
  {Cheng}}, \bibinfo {author} {\bibfnamefont {S.}~\bibnamefont {Wang}},
  \bibinfo {author} {\bibfnamefont {W.-w.}\ \bibnamefont {Xiong}}, \bibinfo
  {author} {\bibfnamefont {L.}~\bibnamefont {Li}}, \bibinfo {author}
  {\bibfnamefont {Z.}~\bibnamefont {An}}, \ and\ \bibinfo {author}
  {\bibfnamefont {W.}~\bibnamefont {Huang}},\ }\href@noop {} {\bibfield
  {journal} {\bibinfo  {journal} {Advanced Materials}\ }\textbf {\bibinfo
  {volume} {29}},\ \bibinfo {pages} {1701244} (\bibinfo {year}
  {2017})}\BibitemShut {NoStop}%
\bibitem [{\citenamefont {Bennett}\ and\ \citenamefont
  {Brassard}(1984)}]{bennett1984quantum}%
  \BibitemOpen
  \bibfield  {author} {\bibinfo {author} {\bibfnamefont {C.~H.}\ \bibnamefont
  {Bennett}}\ and\ \bibinfo {author} {\bibfnamefont {G.}~\bibnamefont
  {Brassard}},\ }in\ \href@noop {} {\emph {\bibinfo {booktitle} {Proceedings of
  IEEE International Conference on Computers, Systems, and Signal Processing,
  Bangalore, India}}}\ (\bibinfo {year} {IEEE, 1984})\ pp.\ \bibinfo {pages}
  {175--179}\BibitemShut {NoStop}%
\bibitem [{\citenamefont {Wang}\ \emph {et~al.}(2010)\citenamefont {Wang},
  \citenamefont {Chen}, \citenamefont {Yin}, \citenamefont {Zhang},
  \citenamefont {Zhang}, \citenamefont {Li}, \citenamefont {Xu}, \citenamefont
  {Zhou}, \citenamefont {Yang}, \citenamefont {Huang}, \citenamefont {Zhang},
  \citenamefont {Li}, \citenamefont {Liu}, \citenamefont {Wang}, \citenamefont
  {Guo},\ and\ \citenamefont {Han}}]{wang2010field}%
  \BibitemOpen
  \bibfield  {author} {\bibinfo {author} {\bibfnamefont {S.}~\bibnamefont
  {Wang}}, \bibinfo {author} {\bibfnamefont {W.}~\bibnamefont {Chen}}, \bibinfo
  {author} {\bibfnamefont {Z.-Q.}\ \bibnamefont {Yin}}, \bibinfo {author}
  {\bibfnamefont {Y.}~\bibnamefont {Zhang}}, \bibinfo {author} {\bibfnamefont
  {T.}~\bibnamefont {Zhang}}, \bibinfo {author} {\bibfnamefont {H.-W.}\
  \bibnamefont {Li}}, \bibinfo {author} {\bibfnamefont {F.-X.}\ \bibnamefont
  {Xu}}, \bibinfo {author} {\bibfnamefont {Z.}~\bibnamefont {Zhou}}, \bibinfo
  {author} {\bibfnamefont {Y.}~\bibnamefont {Yang}}, \bibinfo {author}
  {\bibfnamefont {D.-J.}\ \bibnamefont {Huang}}, \bibinfo {author}
  {\bibfnamefont {L.-J.}\ \bibnamefont {Zhang}}, \bibinfo {author}
  {\bibfnamefont {F.-Y.}\ \bibnamefont {Li}}, \bibinfo {author} {\bibfnamefont
  {D.}~\bibnamefont {Liu}}, \bibinfo {author} {\bibfnamefont {Y.-G.}\
  \bibnamefont {Wang}}, \bibinfo {author} {\bibfnamefont {G.-C.}\ \bibnamefont
  {Guo}}, \ and\ \bibinfo {author} {\bibfnamefont {Z.-F.}\ \bibnamefont
  {Han}},\ }\href@noop {} {\bibfield  {journal} {\bibinfo  {journal} {Optics
  Letters}\ }\textbf {\bibinfo {volume} {35}},\ \bibinfo {pages} {2454}
  (\bibinfo {year} {2010})}\BibitemShut {NoStop}%
\bibitem [{\citenamefont {Sasaki}\ \emph {et~al.}(2011)\citenamefont {Sasaki},
  \citenamefont {Fujiwara}, \citenamefont {Ishizuka}, \citenamefont {Klaus},
  \citenamefont {Wakui}, \citenamefont {Takeoka}, \citenamefont {Miki},
  \citenamefont {Yamashita}, \citenamefont {Wang}, \citenamefont {Tanaka} \emph
  {et~al.}}]{sasaki2011field}%
  \BibitemOpen
  \bibfield  {author} {\bibinfo {author} {\bibfnamefont {M.}~\bibnamefont
  {Sasaki}}, \bibinfo {author} {\bibfnamefont {M.}~\bibnamefont {Fujiwara}},
  \bibinfo {author} {\bibfnamefont {H.}~\bibnamefont {Ishizuka}}, \bibinfo
  {author} {\bibfnamefont {W.}~\bibnamefont {Klaus}}, \bibinfo {author}
  {\bibfnamefont {K.}~\bibnamefont {Wakui}}, \bibinfo {author} {\bibfnamefont
  {M.}~\bibnamefont {Takeoka}}, \bibinfo {author} {\bibfnamefont
  {S.}~\bibnamefont {Miki}}, \bibinfo {author} {\bibfnamefont {T.}~\bibnamefont
  {Yamashita}}, \bibinfo {author} {\bibfnamefont {Z.}~\bibnamefont {Wang}},
  \bibinfo {author} {\bibfnamefont {A.}~\bibnamefont {Tanaka}},  \emph
  {et~al.},\ }\href@noop {} {\bibfield  {journal} {\bibinfo  {journal} {Optics
  express}\ }\textbf {\bibinfo {volume} {19}},\ \bibinfo {pages} {10387}
  (\bibinfo {year} {2011})}\BibitemShut {NoStop}%
\bibitem [{\citenamefont {Wang}\ \emph {et~al.}(2014)\citenamefont {Wang},
  \citenamefont {Chen}, \citenamefont {Yin}, \citenamefont {Li}, \citenamefont
  {He}, \citenamefont {Li}, \citenamefont {Zhou}, \citenamefont {Song},
  \citenamefont {Li}, \citenamefont {Wang}, \citenamefont {Chen}, \citenamefont
  {Han}, \citenamefont {Huang}, \citenamefont {Guo}, \citenamefont {Hao},
  \citenamefont {Li}, \citenamefont {Zhang}, \citenamefont {Liu}, \citenamefont
  {Liang}, \citenamefont {Miao}, \citenamefont {Wu}, \citenamefont {Guo},\ and\
  \citenamefont {Han}}]{wang2014field}%
  \BibitemOpen
  \bibfield  {author} {\bibinfo {author} {\bibfnamefont {S.}~\bibnamefont
  {Wang}}, \bibinfo {author} {\bibfnamefont {W.}~\bibnamefont {Chen}}, \bibinfo
  {author} {\bibfnamefont {Z.-Q.}\ \bibnamefont {Yin}}, \bibinfo {author}
  {\bibfnamefont {H.-W.}\ \bibnamefont {Li}}, \bibinfo {author} {\bibfnamefont
  {D.-Y.}\ \bibnamefont {He}}, \bibinfo {author} {\bibfnamefont {Y.-H.}\
  \bibnamefont {Li}}, \bibinfo {author} {\bibfnamefont {Z.}~\bibnamefont
  {Zhou}}, \bibinfo {author} {\bibfnamefont {X.-T.}\ \bibnamefont {Song}},
  \bibinfo {author} {\bibfnamefont {F.-Y.}\ \bibnamefont {Li}}, \bibinfo
  {author} {\bibfnamefont {D.}~\bibnamefont {Wang}}, \bibinfo {author}
  {\bibfnamefont {H.}~\bibnamefont {Chen}}, \bibinfo {author} {\bibfnamefont
  {Y.-G.}\ \bibnamefont {Han}}, \bibinfo {author} {\bibfnamefont {J.-Z.}\
  \bibnamefont {Huang}}, \bibinfo {author} {\bibfnamefont {J.-F.}\ \bibnamefont
  {Guo}}, \bibinfo {author} {\bibfnamefont {P.-L.}\ \bibnamefont {Hao}},
  \bibinfo {author} {\bibfnamefont {M.}~\bibnamefont {Li}}, \bibinfo {author}
  {\bibfnamefont {C.-M.}\ \bibnamefont {Zhang}}, \bibinfo {author}
  {\bibfnamefont {D.}~\bibnamefont {Liu}}, \bibinfo {author} {\bibfnamefont
  {W.-Y.}\ \bibnamefont {Liang}}, \bibinfo {author} {\bibfnamefont {C.-H.}\
  \bibnamefont {Miao}}, \bibinfo {author} {\bibfnamefont {P.}~\bibnamefont
  {Wu}}, \bibinfo {author} {\bibfnamefont {G.-C.}\ \bibnamefont {Guo}}, \ and\
  \bibinfo {author} {\bibfnamefont {Z.-F.}\ \bibnamefont {Han}},\ }\href@noop
  {} {\bibfield  {journal} {\bibinfo  {journal} {Optics Express}\ }\textbf
  {\bibinfo {volume} {22}},\ \bibinfo {pages} {21739} (\bibinfo {year}
  {2014})}\BibitemShut {NoStop}%
\bibitem [{\citenamefont {Liu}\ \emph {et~al.}(2020)\citenamefont {Liu},
  \citenamefont {Tian}, \citenamefont {Gu}, \citenamefont {Fan}, \citenamefont
  {Ni}, \citenamefont {Yang}, \citenamefont {Zhang}, \citenamefont {Hu},
  \citenamefont {Guo}, \citenamefont {Cao}, \citenamefont {Hu}, \citenamefont
  {Zhao}, \citenamefont {Lu}, \citenamefont {Gong}, \citenamefont {Xie},\ and\
  \citenamefont {Zhu}}]{liu2020drone}%
  \BibitemOpen
  \bibfield  {author} {\bibinfo {author} {\bibfnamefont {H.-Y.}\ \bibnamefont
  {Liu}}, \bibinfo {author} {\bibfnamefont {X.-H.}\ \bibnamefont {Tian}},
  \bibinfo {author} {\bibfnamefont {C.}~\bibnamefont {Gu}}, \bibinfo {author}
  {\bibfnamefont {P.}~\bibnamefont {Fan}}, \bibinfo {author} {\bibfnamefont
  {X.}~\bibnamefont {Ni}}, \bibinfo {author} {\bibfnamefont {R.}~\bibnamefont
  {Yang}}, \bibinfo {author} {\bibfnamefont {J.-N.}\ \bibnamefont {Zhang}},
  \bibinfo {author} {\bibfnamefont {M.}~\bibnamefont {Hu}}, \bibinfo {author}
  {\bibfnamefont {J.}~\bibnamefont {Guo}}, \bibinfo {author} {\bibfnamefont
  {X.}~\bibnamefont {Cao}}, \bibinfo {author} {\bibfnamefont {X.}~\bibnamefont
  {Hu}}, \bibinfo {author} {\bibfnamefont {G.}~\bibnamefont {Zhao}}, \bibinfo
  {author} {\bibfnamefont {Y.-Q.}\ \bibnamefont {Lu}}, \bibinfo {author}
  {\bibfnamefont {Y.-X.}\ \bibnamefont {Gong}}, \bibinfo {author}
  {\bibfnamefont {Z.}~\bibnamefont {Xie}}, \ and\ \bibinfo {author}
  {\bibfnamefont {S.-N.}\ \bibnamefont {Zhu}},\ }\href@noop {} {\bibfield
  {journal} {\bibinfo  {journal} {National Science Review}\ }\textbf {\bibinfo
  {volume} {7}},\ \bibinfo {pages} {921} (\bibinfo {year} {2020})}\BibitemShut
  {NoStop}%
\bibitem [{\citenamefont {Buttler}\ \emph {et~al.}(1998)\citenamefont
  {Buttler}, \citenamefont {Hughes}, \citenamefont {Kwiat}, \citenamefont
  {Lamoreaux}, \citenamefont {Luther}, \citenamefont {Morgan}, \citenamefont
  {Nordholt}, \citenamefont {Peterson},\ and\ \citenamefont
  {Simmons}}]{buttler1998practical}%
  \BibitemOpen
  \bibfield  {author} {\bibinfo {author} {\bibfnamefont {W.~T.}\ \bibnamefont
  {Buttler}}, \bibinfo {author} {\bibfnamefont {R.~J.}\ \bibnamefont {Hughes}},
  \bibinfo {author} {\bibfnamefont {P.~G.}\ \bibnamefont {Kwiat}}, \bibinfo
  {author} {\bibfnamefont {S.~K.}\ \bibnamefont {Lamoreaux}}, \bibinfo {author}
  {\bibfnamefont {G.~G.}\ \bibnamefont {Luther}}, \bibinfo {author}
  {\bibfnamefont {G.~L.}\ \bibnamefont {Morgan}}, \bibinfo {author}
  {\bibfnamefont {J.~E.}\ \bibnamefont {Nordholt}}, \bibinfo {author}
  {\bibfnamefont {C.~G.}\ \bibnamefont {Peterson}}, \ and\ \bibinfo {author}
  {\bibfnamefont {C.~M.}\ \bibnamefont {Simmons}},\ }\href@noop {} {\bibfield
  {journal} {\bibinfo  {journal} {Physical Review Letters}\ }\textbf {\bibinfo
  {volume} {81}},\ \bibinfo {pages} {3283} (\bibinfo {year}
  {1998})}\BibitemShut {NoStop}%
\bibitem [{\citenamefont {Bienfang}\ \emph {et~al.}(2004)\citenamefont
  {Bienfang}, \citenamefont {Gross}, \citenamefont {Mink}, \citenamefont
  {Hershman}, \citenamefont {Nakassis}, \citenamefont {Tang}, \citenamefont
  {Lu}, \citenamefont {Su}, \citenamefont {Clark}, \citenamefont {Williams},
  \citenamefont {Hagley},\ and\ \citenamefont {Wen}}]{bienfang2004quantum}%
  \BibitemOpen
  \bibfield  {author} {\bibinfo {author} {\bibfnamefont {J.~C.}\ \bibnamefont
  {Bienfang}}, \bibinfo {author} {\bibfnamefont {A.~J.}\ \bibnamefont {Gross}},
  \bibinfo {author} {\bibfnamefont {A.}~\bibnamefont {Mink}}, \bibinfo {author}
  {\bibfnamefont {B.~J.}\ \bibnamefont {Hershman}}, \bibinfo {author}
  {\bibfnamefont {A.}~\bibnamefont {Nakassis}}, \bibinfo {author}
  {\bibfnamefont {X.}~\bibnamefont {Tang}}, \bibinfo {author} {\bibfnamefont
  {R.}~\bibnamefont {Lu}}, \bibinfo {author} {\bibfnamefont {D.~H.}\
  \bibnamefont {Su}}, \bibinfo {author} {\bibfnamefont {C.~W.}\ \bibnamefont
  {Clark}}, \bibinfo {author} {\bibfnamefont {C.~J.}\ \bibnamefont {Williams}},
  \bibinfo {author} {\bibfnamefont {E.~W.}\ \bibnamefont {Hagley}}, \ and\
  \bibinfo {author} {\bibfnamefont {J.}~\bibnamefont {Wen}},\ }\href@noop {}
  {\bibfield  {journal} {\bibinfo  {journal} {Optics Express}\ }\textbf
  {\bibinfo {volume} {12}},\ \bibinfo {pages} {2011} (\bibinfo {year}
  {2004})}\BibitemShut {NoStop}%
\bibitem [{\citenamefont {Schmitt-Manderbach}\ \emph
  {et~al.}(2007)\citenamefont {Schmitt-Manderbach}, \citenamefont {Weier},
  \citenamefont {F{\"u}rst}, \citenamefont {Ursin}, \citenamefont
  {Tiefenbacher}, \citenamefont {Scheidl}, \citenamefont {Perdigues},
  \citenamefont {Sodnik}, \citenamefont {Kurtsiefer}, \citenamefont {Rarity},
  \citenamefont {Zeilinger},\ and\ \citenamefont
  {Weinfurter}}]{schmitt2007experimental}%
  \BibitemOpen
  \bibfield  {author} {\bibinfo {author} {\bibfnamefont {T.}~\bibnamefont
  {Schmitt-Manderbach}}, \bibinfo {author} {\bibfnamefont {H.}~\bibnamefont
  {Weier}}, \bibinfo {author} {\bibfnamefont {M.}~\bibnamefont {F{\"u}rst}},
  \bibinfo {author} {\bibfnamefont {R.}~\bibnamefont {Ursin}}, \bibinfo
  {author} {\bibfnamefont {F.}~\bibnamefont {Tiefenbacher}}, \bibinfo {author}
  {\bibfnamefont {T.}~\bibnamefont {Scheidl}}, \bibinfo {author} {\bibfnamefont
  {J.}~\bibnamefont {Perdigues}}, \bibinfo {author} {\bibfnamefont
  {Z.}~\bibnamefont {Sodnik}}, \bibinfo {author} {\bibfnamefont
  {C.}~\bibnamefont {Kurtsiefer}}, \bibinfo {author} {\bibfnamefont {J.~G.}\
  \bibnamefont {Rarity}}, \bibinfo {author} {\bibfnamefont {A.}~\bibnamefont
  {Zeilinger}}, \ and\ \bibinfo {author} {\bibfnamefont {H.}~\bibnamefont
  {Weinfurter}},\ }\href@noop {} {\bibfield  {journal} {\bibinfo  {journal}
  {Physical Review Letters}\ }\textbf {\bibinfo {volume} {98}},\ \bibinfo
  {pages} {010504} (\bibinfo {year} {2007})}\BibitemShut {NoStop}%
\bibitem [{\citenamefont {Tannous}\ \emph {et~al.}(2019)\citenamefont
  {Tannous}, \citenamefont {Ye}, \citenamefont {Jin}, \citenamefont {Kuntz},
  \citenamefont {L{\"u}tkenhaus},\ and\ \citenamefont
  {Jennewein}}]{tannous2019demonstration}%
  \BibitemOpen
  \bibfield  {author} {\bibinfo {author} {\bibfnamefont {R.}~\bibnamefont
  {Tannous}}, \bibinfo {author} {\bibfnamefont {Z.}~\bibnamefont {Ye}},
  \bibinfo {author} {\bibfnamefont {J.}~\bibnamefont {Jin}}, \bibinfo {author}
  {\bibfnamefont {K.~B.}\ \bibnamefont {Kuntz}}, \bibinfo {author}
  {\bibfnamefont {N.}~\bibnamefont {L{\"u}tkenhaus}}, \ and\ \bibinfo {author}
  {\bibfnamefont {T.}~\bibnamefont {Jennewein}},\ }\href@noop {} {\bibfield
  {journal} {\bibinfo  {journal} {Applied Physics Letters}\ }\textbf {\bibinfo
  {volume} {115}},\ \bibinfo {pages} {211103} (\bibinfo {year}
  {2019})}\BibitemShut {NoStop}%
\bibitem [{\citenamefont {Liao}\ \emph {et~al.}(2017)\citenamefont {Liao},
  \citenamefont {Cai}, \citenamefont {Liu}, \citenamefont {Zhang},
  \citenamefont {Li}, \citenamefont {Ren}, \citenamefont {Yin}, \citenamefont
  {Shen}, \citenamefont {Cao}, \citenamefont {Li}, \citenamefont {Li},
  \citenamefont {Chen}, \citenamefont {Sun}, \citenamefont {Jia}, \citenamefont
  {Wu}, \citenamefont {Jiang}, \citenamefont {Wang}, \citenamefont {Huang},
  \citenamefont {Wang}, \citenamefont {Zhou}, \citenamefont {Deng},
  \citenamefont {Xi}, \citenamefont {Ma}, \citenamefont {Hu}, \citenamefont
  {Zhang}, \citenamefont {Chen}, \citenamefont {Liu}, \citenamefont {Wang},
  \citenamefont {Zhu}, \citenamefont {Lu}, \citenamefont {Shu}, \citenamefont
  {Peng}, \citenamefont {Wang},\ and\ \citenamefont {Pan}}]{liao2017satellite}%
  \BibitemOpen
  \bibfield  {author} {\bibinfo {author} {\bibfnamefont {S.-K.}\ \bibnamefont
  {Liao}}, \bibinfo {author} {\bibfnamefont {W.-Q.}\ \bibnamefont {Cai}},
  \bibinfo {author} {\bibfnamefont {W.-Y.}\ \bibnamefont {Liu}}, \bibinfo
  {author} {\bibfnamefont {L.}~\bibnamefont {Zhang}}, \bibinfo {author}
  {\bibfnamefont {Y.}~\bibnamefont {Li}}, \bibinfo {author} {\bibfnamefont
  {J.-G.}\ \bibnamefont {Ren}}, \bibinfo {author} {\bibfnamefont
  {J.}~\bibnamefont {Yin}}, \bibinfo {author} {\bibfnamefont {Q.}~\bibnamefont
  {Shen}}, \bibinfo {author} {\bibfnamefont {Y.}~\bibnamefont {Cao}}, \bibinfo
  {author} {\bibfnamefont {Z.-P.}\ \bibnamefont {Li}}, \bibinfo {author}
  {\bibfnamefont {F.-Z.}\ \bibnamefont {Li}}, \bibinfo {author} {\bibfnamefont
  {X.-W.}\ \bibnamefont {Chen}}, \bibinfo {author} {\bibfnamefont {L.-H.}\
  \bibnamefont {Sun}}, \bibinfo {author} {\bibfnamefont {J.-J.}\ \bibnamefont
  {Jia}}, \bibinfo {author} {\bibfnamefont {J.-C.}\ \bibnamefont {Wu}},
  \bibinfo {author} {\bibfnamefont {X.-J.}\ \bibnamefont {Jiang}}, \bibinfo
  {author} {\bibfnamefont {J.-F.}\ \bibnamefont {Wang}}, \bibinfo {author}
  {\bibfnamefont {Y.-M.}\ \bibnamefont {Huang}}, \bibinfo {author}
  {\bibfnamefont {Q.}~\bibnamefont {Wang}}, \bibinfo {author} {\bibfnamefont
  {Y.-L.}\ \bibnamefont {Zhou}}, \bibinfo {author} {\bibfnamefont
  {L.}~\bibnamefont {Deng}}, \bibinfo {author} {\bibfnamefont {T.}~\bibnamefont
  {Xi}}, \bibinfo {author} {\bibfnamefont {L.}~\bibnamefont {Ma}}, \bibinfo
  {author} {\bibfnamefont {T.}~\bibnamefont {Hu}}, \bibinfo {author}
  {\bibfnamefont {Q.}~\bibnamefont {Zhang}}, \bibinfo {author} {\bibfnamefont
  {Y.-A.}\ \bibnamefont {Chen}}, \bibinfo {author} {\bibfnamefont {N.-L.}\
  \bibnamefont {Liu}}, \bibinfo {author} {\bibfnamefont {X.-B.}\ \bibnamefont
  {Wang}}, \bibinfo {author} {\bibfnamefont {Z.-C.}\ \bibnamefont {Zhu}},
  \bibinfo {author} {\bibfnamefont {C.-Y.}\ \bibnamefont {Lu}}, \bibinfo
  {author} {\bibfnamefont {R.}~\bibnamefont {Shu}}, \bibinfo {author}
  {\bibfnamefont {C.-Z.}\ \bibnamefont {Peng}}, \bibinfo {author}
  {\bibfnamefont {J.-Y.}\ \bibnamefont {Wang}}, \ and\ \bibinfo {author}
  {\bibfnamefont {J.-W.}\ \bibnamefont {Pan}},\ }\href@noop {} {\bibfield
  {journal} {\bibinfo  {journal} {Nature}\ }\textbf {\bibinfo {volume} {549}},\
  \bibinfo {pages} {43} (\bibinfo {year} {2017})}\BibitemShut {NoStop}%
\bibitem [{\citenamefont {Long}\ and\ \citenamefont
  {Liu}(2002)}]{long2002theoretically}%
  \BibitemOpen
  \bibfield  {author} {\bibinfo {author} {\bibfnamefont {G.-L.}\ \bibnamefont
  {Long}}\ and\ \bibinfo {author} {\bibfnamefont {X.-S.}\ \bibnamefont {Liu}},\
  }\href@noop {} {\bibfield  {journal} {\bibinfo  {journal} {Physical Review
  A}\ }\textbf {\bibinfo {volume} {65}},\ \bibinfo {pages} {032302} (\bibinfo
  {year} {2002})},\ \bibinfo {note} {{arXiv preprint quant-ph/0012056,
  2000}}\BibitemShut {NoStop}%
\bibitem [{\citenamefont {Deng}\ \emph {et~al.}(2003)\citenamefont {Deng},
  \citenamefont {Long},\ and\ \citenamefont {Liu}}]{deng2003two}%
  \BibitemOpen
  \bibfield  {author} {\bibinfo {author} {\bibfnamefont {F.-G.}\ \bibnamefont
  {Deng}}, \bibinfo {author} {\bibfnamefont {G.~L.}\ \bibnamefont {Long}}, \
  and\ \bibinfo {author} {\bibfnamefont {X.-S.}\ \bibnamefont {Liu}},\
  }\href@noop {} {\bibfield  {journal} {\bibinfo  {journal} {Physical Review
  A}\ }\textbf {\bibinfo {volume} {68}},\ \bibinfo {pages} {042317} (\bibinfo
  {year} {2003})}\BibitemShut {NoStop}%
\bibitem [{\citenamefont {Deng}\ and\ \citenamefont
  {Long}(2004{\natexlab{a}})}]{deng2004secure}%
  \BibitemOpen
  \bibfield  {author} {\bibinfo {author} {\bibfnamefont {F.-G.}\ \bibnamefont
  {Deng}}\ and\ \bibinfo {author} {\bibfnamefont {G.~L.}\ \bibnamefont
  {Long}},\ }\href@noop {} {\bibfield  {journal} {\bibinfo  {journal} {Physical
  Review A}\ }\textbf {\bibinfo {volume} {69}},\ \bibinfo {pages} {052319}
  (\bibinfo {year} {2004}{\natexlab{a}})}\BibitemShut {NoStop}%
\bibitem [{\citenamefont {Wang}\ \emph {et~al.}(2005)\citenamefont {Wang},
  \citenamefont {Deng}, \citenamefont {Li}, \citenamefont {Liu},\ and\
  \citenamefont {Long}}]{wang2005quantum}%
  \BibitemOpen
  \bibfield  {author} {\bibinfo {author} {\bibfnamefont {C.}~\bibnamefont
  {Wang}}, \bibinfo {author} {\bibfnamefont {F.-G.}\ \bibnamefont {Deng}},
  \bibinfo {author} {\bibfnamefont {Y.-S.}\ \bibnamefont {Li}}, \bibinfo
  {author} {\bibfnamefont {X.-S.}\ \bibnamefont {Liu}}, \ and\ \bibinfo
  {author} {\bibfnamefont {G.~L.}\ \bibnamefont {Long}},\ }\href@noop {}
  {\bibfield  {journal} {\bibinfo  {journal} {Physical Review A}\ }\textbf
  {\bibinfo {volume} {71}},\ \bibinfo {pages} {044305} (\bibinfo {year}
  {2005})}\BibitemShut {NoStop}%
\bibitem [{\citenamefont {Hu}\ \emph {et~al.}(2016)\citenamefont {Hu},
  \citenamefont {Yu}, \citenamefont {Jing}, \citenamefont {Xiao}, \citenamefont
  {Jia}, \citenamefont {Qin},\ and\ \citenamefont {Long}}]{hu2016experimental}%
  \BibitemOpen
  \bibfield  {author} {\bibinfo {author} {\bibfnamefont {J.-Y.}\ \bibnamefont
  {Hu}}, \bibinfo {author} {\bibfnamefont {B.}~\bibnamefont {Yu}}, \bibinfo
  {author} {\bibfnamefont {M.-Y.}\ \bibnamefont {Jing}}, \bibinfo {author}
  {\bibfnamefont {L.-T.}\ \bibnamefont {Xiao}}, \bibinfo {author}
  {\bibfnamefont {S.-T.}\ \bibnamefont {Jia}}, \bibinfo {author} {\bibfnamefont
  {G.-Q.}\ \bibnamefont {Qin}}, \ and\ \bibinfo {author} {\bibfnamefont
  {G.-L.}\ \bibnamefont {Long}},\ }\href@noop {} {\bibfield  {journal}
  {\bibinfo  {journal} {Light: Science \& Applications}\ }\textbf {\bibinfo
  {volume} {5}},\ \bibinfo {pages} {e16144} (\bibinfo {year}
  {2016})}\BibitemShut {NoStop}%
\bibitem [{\citenamefont {Zhang}\ \emph {et~al.}(2017)\citenamefont {Zhang},
  \citenamefont {Ding}, \citenamefont {Sheng}, \citenamefont {Zhou},
  \citenamefont {Shi},\ and\ \citenamefont {Guo}}]{zhang2017quantum}%
  \BibitemOpen
  \bibfield  {author} {\bibinfo {author} {\bibfnamefont {W.}~\bibnamefont
  {Zhang}}, \bibinfo {author} {\bibfnamefont {D.-S.}\ \bibnamefont {Ding}},
  \bibinfo {author} {\bibfnamefont {Y.-B.}\ \bibnamefont {Sheng}}, \bibinfo
  {author} {\bibfnamefont {L.}~\bibnamefont {Zhou}}, \bibinfo {author}
  {\bibfnamefont {B.-S.}\ \bibnamefont {Shi}}, \ and\ \bibinfo {author}
  {\bibfnamefont {G.-C.}\ \bibnamefont {Guo}},\ }\href@noop {} {\bibfield
  {journal} {\bibinfo  {journal} {Physical Review Letters}\ }\textbf {\bibinfo
  {volume} {118}},\ \bibinfo {pages} {220501} (\bibinfo {year}
  {2017})}\BibitemShut {NoStop}%
\bibitem [{\citenamefont {Zhu}\ \emph {et~al.}(2017)\citenamefont {Zhu},
  \citenamefont {Zhang}, \citenamefont {Sheng},\ and\ \citenamefont
  {Huang}}]{zhu2017experimental}%
  \BibitemOpen
  \bibfield  {author} {\bibinfo {author} {\bibfnamefont {F.}~\bibnamefont
  {Zhu}}, \bibinfo {author} {\bibfnamefont {W.}~\bibnamefont {Zhang}}, \bibinfo
  {author} {\bibfnamefont {Y.}~\bibnamefont {Sheng}}, \ and\ \bibinfo {author}
  {\bibfnamefont {Y.}~\bibnamefont {Huang}},\ }\href@noop {} {\bibfield
  {journal} {\bibinfo  {journal} {Science Bulletin}\ }\textbf {\bibinfo
  {volume} {62}},\ \bibinfo {pages} {1519} (\bibinfo {year}
  {2017})}\BibitemShut {NoStop}%
\bibitem [{\citenamefont {Qi}\ \emph {et~al.}(2019)\citenamefont {Qi},
  \citenamefont {Sun}, \citenamefont {Lin}, \citenamefont {Niu}, \citenamefont
  {Hao}, \citenamefont {Song}, \citenamefont {Huang}, \citenamefont {Gao},
  \citenamefont {Yin},\ and\ \citenamefont {Long}}]{qi2019implementation}%
  \BibitemOpen
  \bibfield  {author} {\bibinfo {author} {\bibfnamefont {R.}~\bibnamefont
  {Qi}}, \bibinfo {author} {\bibfnamefont {Z.}~\bibnamefont {Sun}}, \bibinfo
  {author} {\bibfnamefont {Z.}~\bibnamefont {Lin}}, \bibinfo {author}
  {\bibfnamefont {P.}~\bibnamefont {Niu}}, \bibinfo {author} {\bibfnamefont
  {W.}~\bibnamefont {Hao}}, \bibinfo {author} {\bibfnamefont {L.}~\bibnamefont
  {Song}}, \bibinfo {author} {\bibfnamefont {Q.}~\bibnamefont {Huang}},
  \bibinfo {author} {\bibfnamefont {J.}~\bibnamefont {Gao}}, \bibinfo {author}
  {\bibfnamefont {L.}~\bibnamefont {Yin}}, \ and\ \bibinfo {author}
  {\bibfnamefont {G.-L.}\ \bibnamefont {Long}},\ }\href@noop {} {\bibfield
  {journal} {\bibinfo  {journal} {Light: Science \& Applications}\ }\textbf
  {\bibinfo {volume} {8}},\ \bibinfo {pages} {22} (\bibinfo {year}
  {2019})}\BibitemShut {NoStop}%
\bibitem [{\citenamefont {Marino}\ and\ \citenamefont
  {Stroud~Jr}(2006)}]{marino2006deterministic}%
  \BibitemOpen
  \bibfield  {author} {\bibinfo {author} {\bibfnamefont {A.~M.}\ \bibnamefont
  {Marino}}\ and\ \bibinfo {author} {\bibfnamefont {C.}~\bibnamefont
  {Stroud~Jr}},\ }\href@noop {} {\bibfield  {journal} {\bibinfo  {journal}
  {Physical Review A}\ }\textbf {\bibinfo {volume} {74}},\ \bibinfo {pages}
  {022315} (\bibinfo {year} {2006})}\BibitemShut {NoStop}%
\bibitem [{\citenamefont {Shapiro}\ \emph {et~al.}(2014)\citenamefont
  {Shapiro}, \citenamefont {Zhang},\ and\ \citenamefont
  {Wong}}]{shapiro2014secure}%
  \BibitemOpen
  \bibfield  {author} {\bibinfo {author} {\bibfnamefont {J.~H.}\ \bibnamefont
  {Shapiro}}, \bibinfo {author} {\bibfnamefont {Z.}~\bibnamefont {Zhang}}, \
  and\ \bibinfo {author} {\bibfnamefont {F.~N.}\ \bibnamefont {Wong}},\
  }\href@noop {} {\bibfield  {journal} {\bibinfo  {journal} {Quantum
  Information Processing}\ }\textbf {\bibinfo {volume} {13}},\ \bibinfo {pages}
  {2171} (\bibinfo {year} {2014})}\BibitemShut {NoStop}%
\bibitem [{\citenamefont {Pirandola}\ \emph {et~al.}(2008)\citenamefont
  {Pirandola}, \citenamefont {Braunstein}, \citenamefont {Mancini},\ and\
  \citenamefont {Lloyd}}]{pirandola2008quantum}%
  \BibitemOpen
  \bibfield  {author} {\bibinfo {author} {\bibfnamefont {S.}~\bibnamefont
  {Pirandola}}, \bibinfo {author} {\bibfnamefont {S.~L.}\ \bibnamefont
  {Braunstein}}, \bibinfo {author} {\bibfnamefont {S.}~\bibnamefont {Mancini}},
  \ and\ \bibinfo {author} {\bibfnamefont {S.}~\bibnamefont {Lloyd}},\
  }\href@noop {} {\bibfield  {journal} {\bibinfo  {journal} {Europhysics
  Letters}\ }\textbf {\bibinfo {volume} {84}},\ \bibinfo {pages} {20013}
  (\bibinfo {year} {2008})}\BibitemShut {NoStop}%
\bibitem [{\citenamefont {Pirandola}\ \emph {et~al.}(2009)\citenamefont
  {Pirandola}, \citenamefont {Braunstein}, \citenamefont {Lloyd},\ and\
  \citenamefont {Mancini}}]{pirandola2009confidential}%
  \BibitemOpen
  \bibfield  {author} {\bibinfo {author} {\bibfnamefont {S.}~\bibnamefont
  {Pirandola}}, \bibinfo {author} {\bibfnamefont {S.~L.}\ \bibnamefont
  {Braunstein}}, \bibinfo {author} {\bibfnamefont {S.}~\bibnamefont {Lloyd}}, \
  and\ \bibinfo {author} {\bibfnamefont {S.}~\bibnamefont {Mancini}},\
  }\href@noop {} {\bibfield  {journal} {\bibinfo  {journal} {IEEE Journal of
  Selected Topics in Quantum Electronics}\ }\textbf {\bibinfo {volume} {15}},\
  \bibinfo {pages} {1570} (\bibinfo {year} {2009})}\BibitemShut {NoStop}%
\bibitem [{\citenamefont {Lum}\ \emph {et~al.}(2016)\citenamefont {Lum},
  \citenamefont {Howell}, \citenamefont {Allman}, \citenamefont {Gerrits},
  \citenamefont {Verma}, \citenamefont {Nam}, \citenamefont {Lupo},\ and\
  \citenamefont {Lloyd}}]{lum2016quantum}%
  \BibitemOpen
  \bibfield  {author} {\bibinfo {author} {\bibfnamefont {D.~J.}\ \bibnamefont
  {Lum}}, \bibinfo {author} {\bibfnamefont {J.~C.}\ \bibnamefont {Howell}},
  \bibinfo {author} {\bibfnamefont {M.~S.}\ \bibnamefont {Allman}}, \bibinfo
  {author} {\bibfnamefont {T.}~\bibnamefont {Gerrits}}, \bibinfo {author}
  {\bibfnamefont {V.~B.}\ \bibnamefont {Verma}}, \bibinfo {author}
  {\bibfnamefont {S.~W.}\ \bibnamefont {Nam}}, \bibinfo {author} {\bibfnamefont
  {C.}~\bibnamefont {Lupo}}, \ and\ \bibinfo {author} {\bibfnamefont
  {S.}~\bibnamefont {Lloyd}},\ }\href@noop {} {\bibfield  {journal} {\bibinfo
  {journal} {Physical Review A}\ }\textbf {\bibinfo {volume} {94}},\ \bibinfo
  {pages} {022315} (\bibinfo {year} {2016})}\BibitemShut {NoStop}%
\bibitem [{\citenamefont {Zhou}\ \emph {et~al.}(2020)\citenamefont {Zhou},
  \citenamefont {Sheng}, \citenamefont {Niu}, \citenamefont {Yin},
  \citenamefont {Long},\ and\ \citenamefont {Hanzo}}]{zhou2020measurement}%
  \BibitemOpen
  \bibfield  {author} {\bibinfo {author} {\bibfnamefont {Z.-R.}\ \bibnamefont
  {Zhou}}, \bibinfo {author} {\bibfnamefont {Y.-B.}\ \bibnamefont {Sheng}},
  \bibinfo {author} {\bibfnamefont {P.-H.}\ \bibnamefont {Niu}}, \bibinfo
  {author} {\bibfnamefont {L.-G.}\ \bibnamefont {Yin}}, \bibinfo {author}
  {\bibfnamefont {G.-L.}\ \bibnamefont {Long}}, \ and\ \bibinfo {author}
  {\bibfnamefont {L.}~\bibnamefont {Hanzo}},\ }\href@noop {} {\bibfield
  {journal} {\bibinfo  {journal} {Science China Physics, Mechanics \&
  Astronomy}\ }\textbf {\bibinfo {volume} {63}},\ \bibinfo {pages} {230362}
  (\bibinfo {year} {2020})}\BibitemShut {NoStop}%
\bibitem [{\citenamefont {Niu}\ \emph {et~al.}(2018)\citenamefont {Niu},
  \citenamefont {Zhou}, \citenamefont {Lin}, \citenamefont {Sheng},
  \citenamefont {Yin},\ and\ \citenamefont {Long}}]{niu2018measurement}%
  \BibitemOpen
  \bibfield  {author} {\bibinfo {author} {\bibfnamefont {P.-H.}\ \bibnamefont
  {Niu}}, \bibinfo {author} {\bibfnamefont {Z.-R.}\ \bibnamefont {Zhou}},
  \bibinfo {author} {\bibfnamefont {Z.-S.}\ \bibnamefont {Lin}}, \bibinfo
  {author} {\bibfnamefont {Y.-B.}\ \bibnamefont {Sheng}}, \bibinfo {author}
  {\bibfnamefont {L.-G.}\ \bibnamefont {Yin}}, \ and\ \bibinfo {author}
  {\bibfnamefont {G.-L.}\ \bibnamefont {Long}},\ }\href@noop {} {\bibfield
  {journal} {\bibinfo  {journal} {Science Bulletin}\ }\textbf {\bibinfo
  {volume} {63}},\ \bibinfo {pages} {1345} (\bibinfo {year}
  {2018})}\BibitemShut {NoStop}%
\bibitem [{\citenamefont {Gao}\ \emph {et~al.}(2019)\citenamefont {Gao},
  \citenamefont {Li},\ and\ \citenamefont {Li}}]{gao2019long}%
  \BibitemOpen
  \bibfield  {author} {\bibinfo {author} {\bibfnamefont {Z.}~\bibnamefont
  {Gao}}, \bibinfo {author} {\bibfnamefont {T.}~\bibnamefont {Li}}, \ and\
  \bibinfo {author} {\bibfnamefont {Z.}~\bibnamefont {Li}},\ }\href@noop {}
  {\bibfield  {journal} {\bibinfo  {journal} {Europhysics Letters}\ }\textbf
  {\bibinfo {volume} {125}},\ \bibinfo {pages} {40004} (\bibinfo {year}
  {2019})}\BibitemShut {NoStop}%
\bibitem [{\citenamefont {Zhou}\ \emph
  {et~al.}(2019{\natexlab{a}})\citenamefont {Zhou}, \citenamefont {Sheng},\
  and\ \citenamefont {Long}}]{zhou2019device}%
  \BibitemOpen
  \bibfield  {author} {\bibinfo {author} {\bibfnamefont {L.}~\bibnamefont
  {Zhou}}, \bibinfo {author} {\bibfnamefont {Y.-B.}\ \bibnamefont {Sheng}}, \
  and\ \bibinfo {author} {\bibfnamefont {G.-L.}\ \bibnamefont {Long}},\
  }\href@noop {} {\bibfield  {journal} {\bibinfo  {journal} {Science Bulletin}\
  } (\bibinfo {year} {2019}{\natexlab{a}})}\BibitemShut {NoStop}%
\bibitem [{\citenamefont {Gottesman}\ \emph {et~al.}(2004)\citenamefont
  {Gottesman}, \citenamefont {Lo}, \citenamefont {L{\"u}tkenhaus},\ and\
  \citenamefont {Preskill}}]{gottesman2004security}%
  \BibitemOpen
  \bibfield  {author} {\bibinfo {author} {\bibfnamefont {D.}~\bibnamefont
  {Gottesman}}, \bibinfo {author} {\bibfnamefont {H.-K.}\ \bibnamefont {Lo}},
  \bibinfo {author} {\bibfnamefont {N.}~\bibnamefont {L{\"u}tkenhaus}}, \ and\
  \bibinfo {author} {\bibfnamefont {J.}~\bibnamefont {Preskill}},\ }\href@noop
  {} {\bibfield  {journal} {\bibinfo  {journal} {Quantum Information \&
  Computation}\ }\textbf {\bibinfo {volume} {4}},\ \bibinfo {pages} {325}
  (\bibinfo {year} {2004})}\BibitemShut {NoStop}%
\bibitem [{\citenamefont {Hwang}(2003)}]{hwang2003quantum}%
  \BibitemOpen
  \bibfield  {author} {\bibinfo {author} {\bibfnamefont {W.-Y.}\ \bibnamefont
  {Hwang}},\ }\href@noop {} {\bibfield  {journal} {\bibinfo  {journal}
  {Physical Review Letters}\ }\textbf {\bibinfo {volume} {91}},\ \bibinfo
  {pages} {057901} (\bibinfo {year} {2003})}\BibitemShut {NoStop}%
\bibitem [{\citenamefont {Wang}(2005)}]{wang2005beating}%
  \BibitemOpen
  \bibfield  {author} {\bibinfo {author} {\bibfnamefont {X.-B.}\ \bibnamefont
  {Wang}},\ }\href@noop {} {\bibfield  {journal} {\bibinfo  {journal} {Physical
  Review Letters}\ }\textbf {\bibinfo {volume} {94}},\ \bibinfo {pages}
  {230503} (\bibinfo {year} {2005})}\BibitemShut {NoStop}%
\bibitem [{\citenamefont {Lo}\ \emph {et~al.}(2005)\citenamefont {Lo},
  \citenamefont {Ma},\ and\ \citenamefont {Chen}}]{lo2005decoy}%
  \BibitemOpen
  \bibfield  {author} {\bibinfo {author} {\bibfnamefont {H.-K.}\ \bibnamefont
  {Lo}}, \bibinfo {author} {\bibfnamefont {X.}~\bibnamefont {Ma}}, \ and\
  \bibinfo {author} {\bibfnamefont {K.}~\bibnamefont {Chen}},\ }\href@noop {}
  {\bibfield  {journal} {\bibinfo  {journal} {Physical Review Letters}\
  }\textbf {\bibinfo {volume} {94}},\ \bibinfo {pages} {230504} (\bibinfo
  {year} {2005})}\BibitemShut {NoStop}%
\bibitem [{\citenamefont {Deng}\ and\ \citenamefont
  {Long}(2004{\natexlab{b}})}]{deng2004bidirectional}%
  \BibitemOpen
  \bibfield  {author} {\bibinfo {author} {\bibfnamefont {F.-G.}\ \bibnamefont
  {Deng}}\ and\ \bibinfo {author} {\bibfnamefont {G.~L.}\ \bibnamefont
  {Long}},\ }\href@noop {} {\bibfield  {journal} {\bibinfo  {journal} {Physical
  Review A}\ }\textbf {\bibinfo {volume} {70}},\ \bibinfo {pages} {012311}
  (\bibinfo {year} {2004}{\natexlab{b}})}\BibitemShut {NoStop}%
\bibitem [{\citenamefont {Lucamarini}\ and\ \citenamefont
  {Mancini}(2005)}]{lucamarini2005secure}%
  \BibitemOpen
  \bibfield  {author} {\bibinfo {author} {\bibfnamefont {M.}~\bibnamefont
  {Lucamarini}}\ and\ \bibinfo {author} {\bibfnamefont {S.}~\bibnamefont
  {Mancini}},\ }\href@noop {} {\bibfield  {journal} {\bibinfo  {journal}
  {Physical Review Letters}\ }\textbf {\bibinfo {volume} {94}},\ \bibinfo
  {pages} {140501} (\bibinfo {year} {2005})}\BibitemShut {NoStop}%
\bibitem [{\citenamefont {Lu}(2019)}]{lu2019ambiguous}%
  \BibitemOpen
  \bibfield  {author} {\bibinfo {author} {\bibfnamefont {H.}~\bibnamefont
  {Lu}},\ }\href@noop {} {\bibfield  {journal} {\bibinfo  {journal} {JOSA B}\
  }\textbf {\bibinfo {volume} {36}},\ \bibinfo {pages} {B26} (\bibinfo {year}
  {2019})}\BibitemShut {NoStop}%
\bibitem [{\citenamefont {Zhou}\ \emph
  {et~al.}(2019{\natexlab{b}})\citenamefont {Zhou}, \citenamefont {Valivarthi},
  \citenamefont {John},\ and\ \citenamefont {Tittel}}]{zhou2019practical}%
  \BibitemOpen
  \bibfield  {author} {\bibinfo {author} {\bibfnamefont {Q.}~\bibnamefont
  {Zhou}}, \bibinfo {author} {\bibfnamefont {R.}~\bibnamefont {Valivarthi}},
  \bibinfo {author} {\bibfnamefont {C.}~\bibnamefont {John}}, \ and\ \bibinfo
  {author} {\bibfnamefont {W.}~\bibnamefont {Tittel}},\ }\href@noop {}
  {\bibfield  {journal} {\bibinfo  {journal} {Quantum Engineering}\ }\textbf
  {\bibinfo {volume} {1}},\ \bibinfo {pages} {e8} (\bibinfo {year}
  {2019}{\natexlab{b}})}\BibitemShut {NoStop}%
\bibitem [{\citenamefont {Muller}\ \emph {et~al.}(1997)\citenamefont {Muller},
  \citenamefont {Herzog}, \citenamefont {Huttner}, \citenamefont {Tittel},
  \citenamefont {Zbinden},\ and\ \citenamefont {Gisin}}]{muller1997plug}%
  \BibitemOpen
  \bibfield  {author} {\bibinfo {author} {\bibfnamefont {A.}~\bibnamefont
  {Muller}}, \bibinfo {author} {\bibfnamefont {T.}~\bibnamefont {Herzog}},
  \bibinfo {author} {\bibfnamefont {B.}~\bibnamefont {Huttner}}, \bibinfo
  {author} {\bibfnamefont {W.}~\bibnamefont {Tittel}}, \bibinfo {author}
  {\bibfnamefont {H.}~\bibnamefont {Zbinden}}, \ and\ \bibinfo {author}
  {\bibfnamefont {N.}~\bibnamefont {Gisin}},\ }\href@noop {} {\bibfield
  {journal} {\bibinfo  {journal} {Applied Physics Letters}\ }\textbf {\bibinfo
  {volume} {70}},\ \bibinfo {pages} {793} (\bibinfo {year} {1997})}\BibitemShut
  {NoStop}%
\bibitem [{\citenamefont {Wang}\ \emph {et~al.}(2018)\citenamefont {Wang},
  \citenamefont {Chen}, \citenamefont {Yin}, \citenamefont {He}, \citenamefont
  {Hui}, \citenamefont {Hao}, \citenamefont {Fan-Yuan}, \citenamefont {Wang},
  \citenamefont {Zhang}, \citenamefont {Kuang}, \citenamefont {Liu},
  \citenamefont {Zhou}, \citenamefont {Wang}, \citenamefont {Guo},\ and\
  \citenamefont {Han}}]{wang2018practical}%
  \BibitemOpen
  \bibfield  {author} {\bibinfo {author} {\bibfnamefont {S.}~\bibnamefont
  {Wang}}, \bibinfo {author} {\bibfnamefont {W.}~\bibnamefont {Chen}}, \bibinfo
  {author} {\bibfnamefont {Z.-Q.}\ \bibnamefont {Yin}}, \bibinfo {author}
  {\bibfnamefont {D.-Y.}\ \bibnamefont {He}}, \bibinfo {author} {\bibfnamefont
  {C.}~\bibnamefont {Hui}}, \bibinfo {author} {\bibfnamefont {P.-L.}\
  \bibnamefont {Hao}}, \bibinfo {author} {\bibfnamefont {G.-J.}\ \bibnamefont
  {Fan-Yuan}}, \bibinfo {author} {\bibfnamefont {C.}~\bibnamefont {Wang}},
  \bibinfo {author} {\bibfnamefont {L.-J.}\ \bibnamefont {Zhang}}, \bibinfo
  {author} {\bibfnamefont {J.}~\bibnamefont {Kuang}}, \bibinfo {author}
  {\bibfnamefont {S.-F.}\ \bibnamefont {Liu}}, \bibinfo {author} {\bibfnamefont
  {Z.}~\bibnamefont {Zhou}}, \bibinfo {author} {\bibfnamefont {Y.-G.}\
  \bibnamefont {Wang}}, \bibinfo {author} {\bibfnamefont {G.-C.}\ \bibnamefont
  {Guo}}, \ and\ \bibinfo {author} {\bibfnamefont {Z.-F.}\ \bibnamefont
  {Han}},\ }\href@noop {} {\bibfield  {journal} {\bibinfo  {journal} {Optics
  Letters}\ }\textbf {\bibinfo {volume} {43}},\ \bibinfo {pages} {2030}
  (\bibinfo {year} {2018})}\BibitemShut {NoStop}%
\bibitem [{\citenamefont {Thangaraj}\ \emph {et~al.}(2007)\citenamefont
  {Thangaraj}, \citenamefont {Dihidar}, \citenamefont {Calderbank},
  \citenamefont {McLaughlin},\ and\ \citenamefont
  {Merolla}}]{thangaraj2007applications}%
  \BibitemOpen
  \bibfield  {author} {\bibinfo {author} {\bibfnamefont {A.}~\bibnamefont
  {Thangaraj}}, \bibinfo {author} {\bibfnamefont {S.}~\bibnamefont {Dihidar}},
  \bibinfo {author} {\bibfnamefont {A.~R.}\ \bibnamefont {Calderbank}},
  \bibinfo {author} {\bibfnamefont {S.~W.}\ \bibnamefont {McLaughlin}}, \ and\
  \bibinfo {author} {\bibfnamefont {J.-M.}\ \bibnamefont {Merolla}},\
  }\href@noop {} {\bibfield  {journal} {\bibinfo  {journal} {IEEE Trans. Inf.
  Theory}\ }\textbf {\bibinfo {volume} {53}},\ \bibinfo {pages} {2933}
  (\bibinfo {year} {2007})}\BibitemShut {NoStop}%
\bibitem [{\citenamefont {Wyner}(1975)}]{wyner1975wire}%
  \BibitemOpen
  \bibfield  {author} {\bibinfo {author} {\bibfnamefont {A.~D.}\ \bibnamefont
  {Wyner}},\ }\href@noop {} {\bibfield  {journal} {\bibinfo  {journal} {Bell
  System Technical Journal}\ }\textbf {\bibinfo {volume} {54}},\ \bibinfo
  {pages} {1355} (\bibinfo {year} {1975})}\BibitemShut {NoStop}%
\bibitem [{\citenamefont {Tomamichel}\ \emph {et~al.}(2012)\citenamefont
  {Tomamichel}, \citenamefont {Lim}, \citenamefont {Gisin},\ and\ \citenamefont
  {Renner}}]{tomamichel2012tight}%
  \BibitemOpen
  \bibfield  {author} {\bibinfo {author} {\bibfnamefont {M.}~\bibnamefont
  {Tomamichel}}, \bibinfo {author} {\bibfnamefont {C.~C.~W.}\ \bibnamefont
  {Lim}}, \bibinfo {author} {\bibfnamefont {N.}~\bibnamefont {Gisin}}, \ and\
  \bibinfo {author} {\bibfnamefont {R.}~\bibnamefont {Renner}},\ }\href@noop {}
  {\bibfield  {journal} {\bibinfo  {journal} {Nature communications}\ }\textbf
  {\bibinfo {volume} {3}},\ \bibinfo {pages} {1} (\bibinfo {year}
  {2012})}\BibitemShut {NoStop}%
\bibitem [{\citenamefont {Lu}\ \emph {et~al.}(2011)\citenamefont {Lu},
  \citenamefont {Fung}, \citenamefont {Ma},\ and\ \citenamefont
  {Cai}}]{lu2011unconditional}%
  \BibitemOpen
  \bibfield  {author} {\bibinfo {author} {\bibfnamefont {H.}~\bibnamefont
  {Lu}}, \bibinfo {author} {\bibfnamefont {C.-H.~F.}\ \bibnamefont {Fung}},
  \bibinfo {author} {\bibfnamefont {X.}~\bibnamefont {Ma}}, \ and\ \bibinfo
  {author} {\bibfnamefont {Q.-y.}\ \bibnamefont {Cai}},\ }\href@noop {}
  {\bibfield  {journal} {\bibinfo  {journal} {Physical Review A}\ }\textbf
  {\bibinfo {volume} {84}},\ \bibinfo {pages} {042344} (\bibinfo {year}
  {2011})}\BibitemShut {NoStop}%
\bibitem [{\citenamefont {Henao}\ and\ \citenamefont
  {Serra}(2015)}]{henao2015practical}%
  \BibitemOpen
  \bibfield  {author} {\bibinfo {author} {\bibfnamefont {C.~I.}\ \bibnamefont
  {Henao}}\ and\ \bibinfo {author} {\bibfnamefont {R.~M.}\ \bibnamefont
  {Serra}},\ }\href@noop {} {\bibfield  {journal} {\bibinfo  {journal}
  {Physical Review A}\ }\textbf {\bibinfo {volume} {92}},\ \bibinfo {pages}
  {052317} (\bibinfo {year} {2015})}\BibitemShut {NoStop}%
\bibitem [{\citenamefont {Wu}\ \emph {et~al.}(2019)\citenamefont {Wu},
  \citenamefont {Lin}, \citenamefont {Yin},\ and\ \citenamefont
  {Long}}]{wu2019security}%
  \BibitemOpen
  \bibfield  {author} {\bibinfo {author} {\bibfnamefont {J.}~\bibnamefont
  {Wu}}, \bibinfo {author} {\bibfnamefont {Z.}~\bibnamefont {Lin}}, \bibinfo
  {author} {\bibfnamefont {L.}~\bibnamefont {Yin}}, \ and\ \bibinfo {author}
  {\bibfnamefont {G.-L.}\ \bibnamefont {Long}},\ }\href@noop {} {\bibfield
  {journal} {\bibinfo  {journal} {Quantum Engineering}\ }\textbf {\bibinfo
  {volume} {1}},\ \bibinfo {pages} {e26} (\bibinfo {year} {2019})}\BibitemShut
  {NoStop}%
\bibitem [{\citenamefont {Feng}\ \emph {et~al.}(2004)\citenamefont {Feng},
  \citenamefont {Duan},\ and\ \citenamefont {Ying}}]{feng2004unambiguous}%
  \BibitemOpen
  \bibfield  {author} {\bibinfo {author} {\bibfnamefont {Y.}~\bibnamefont
  {Feng}}, \bibinfo {author} {\bibfnamefont {R.}~\bibnamefont {Duan}}, \ and\
  \bibinfo {author} {\bibfnamefont {M.}~\bibnamefont {Ying}},\ }\href@noop {}
  {\bibfield  {journal} {\bibinfo  {journal} {Physical Review A}\ }\textbf
  {\bibinfo {volume} {70}},\ \bibinfo {pages} {012308} (\bibinfo {year}
  {2004})}\BibitemShut {NoStop}%
\bibitem [{\citenamefont {Holevo}(1973)}]{holevo1973bounds}%
  \BibitemOpen
  \bibfield  {author} {\bibinfo {author} {\bibfnamefont {A.~S.}\ \bibnamefont
  {Holevo}},\ }\href@noop {} {\bibfield  {journal} {\bibinfo  {journal}
  {Problems of Information Transmission}\ }\textbf {\bibinfo {volume} {9}},\
  \bibinfo {pages} {177} (\bibinfo {year} {1973})}\BibitemShut {NoStop}%
\bibitem [{\citenamefont {Jozsa}\ and\ \citenamefont
  {Schlienz}(2000)}]{jozsa2000distinguishability}%
  \BibitemOpen
  \bibfield  {author} {\bibinfo {author} {\bibfnamefont {R.}~\bibnamefont
  {Jozsa}}\ and\ \bibinfo {author} {\bibfnamefont {J.}~\bibnamefont
  {Schlienz}},\ }\href@noop {} {\bibfield  {journal} {\bibinfo  {journal}
  {Physical Review A}\ }\textbf {\bibinfo {volume} {62}},\ \bibinfo {pages}
  {012301} (\bibinfo {year} {2000})}\BibitemShut {NoStop}%
\bibitem [{\citenamefont {Krawec}(2017)}]{krawec2017quantum}%
  \BibitemOpen
  \bibfield  {author} {\bibinfo {author} {\bibfnamefont {W.~O.}\ \bibnamefont
  {Krawec}},\ }\href@noop {} {\bibfield  {journal} {\bibinfo  {journal}
  {Quantum Information \& Computation}\ }\textbf {\bibinfo {volume} {17}},\
  \bibinfo {pages} {209} (\bibinfo {year} {2017})}\BibitemShut {NoStop}%
\bibitem [{\citenamefont {Christandl}\ \emph {et~al.}(2004)\citenamefont
  {Christandl}, \citenamefont {Renner},\ and\ \citenamefont
  {Ekert}}]{christandl2004generic}%
  \BibitemOpen
  \bibfield  {author} {\bibinfo {author} {\bibfnamefont {M.}~\bibnamefont
  {Christandl}}, \bibinfo {author} {\bibfnamefont {R.}~\bibnamefont {Renner}},
  \ and\ \bibinfo {author} {\bibfnamefont {A.}~\bibnamefont {Ekert}},\
  }\href@noop {} {\bibfield  {journal} {\bibinfo  {journal} {arXiv preprint
  quant-ph/0402131}\ } (\bibinfo {year} {2004})}\BibitemShut {NoStop}%
\bibitem [{\citenamefont {Scarani}\ \emph {et~al.}(2009)\citenamefont
  {Scarani}, \citenamefont {Bechmann-Pasquinucci}, \citenamefont {Cerf},
  \citenamefont {Du{\v{s}}ek}, \citenamefont {L{\"u}tkenhaus},\ and\
  \citenamefont {Peev}}]{scarani2009security}%
  \BibitemOpen
  \bibfield  {author} {\bibinfo {author} {\bibfnamefont {V.}~\bibnamefont
  {Scarani}}, \bibinfo {author} {\bibfnamefont {H.}~\bibnamefont
  {Bechmann-Pasquinucci}}, \bibinfo {author} {\bibfnamefont {N.~J.}\
  \bibnamefont {Cerf}}, \bibinfo {author} {\bibfnamefont {M.}~\bibnamefont
  {Du{\v{s}}ek}}, \bibinfo {author} {\bibfnamefont {N.}~\bibnamefont
  {L{\"u}tkenhaus}}, \ and\ \bibinfo {author} {\bibfnamefont {M.}~\bibnamefont
  {Peev}},\ }\href@noop {} {\bibfield  {journal} {\bibinfo  {journal} {Reviews
  of Modern Physics}\ }\textbf {\bibinfo {volume} {81}},\ \bibinfo {pages}
  {1301} (\bibinfo {year} {2009})}\BibitemShut {NoStop}%
\bibitem [{\citenamefont {Ma}\ \emph {et~al.}(2005)\citenamefont {Ma},
  \citenamefont {Qi}, \citenamefont {Zhao},\ and\ \citenamefont
  {Lo}}]{ma2005practical}%
  \BibitemOpen
  \bibfield  {author} {\bibinfo {author} {\bibfnamefont {X.}~\bibnamefont
  {Ma}}, \bibinfo {author} {\bibfnamefont {B.}~\bibnamefont {Qi}}, \bibinfo
  {author} {\bibfnamefont {Y.}~\bibnamefont {Zhao}}, \ and\ \bibinfo {author}
  {\bibfnamefont {H.-K.}\ \bibnamefont {Lo}},\ }\href@noop {} {\bibfield
  {journal} {\bibinfo  {journal} {Physical Review A}\ }\textbf {\bibinfo
  {volume} {72}},\ \bibinfo {pages} {012326} (\bibinfo {year}
  {2005})}\BibitemShut {NoStop}%
\bibitem [{\citenamefont {Ma}\ and\ \citenamefont {Lo}(2008)}]{ma2008quantum}%
  \BibitemOpen
  \bibfield  {author} {\bibinfo {author} {\bibfnamefont {X.}~\bibnamefont
  {Ma}}\ and\ \bibinfo {author} {\bibfnamefont {H.-K.}\ \bibnamefont {Lo}},\
  }\href@noop {} {\bibfield  {journal} {\bibinfo  {journal} {New Journal of
  Physics}\ }\textbf {\bibinfo {volume} {10}},\ \bibinfo {pages} {073018}
  (\bibinfo {year} {2008})}\BibitemShut {NoStop}%
\bibitem [{\citenamefont {MacKay}(2003)}]{mackay2003information}%
  \BibitemOpen
  \bibfield  {author} {\bibinfo {author} {\bibfnamefont {D.~J.~C.}\
  \bibnamefont {MacKay}},\ }\href@noop {} {\emph {\bibinfo {title} {Information
  theory, inference and learning algorithms}}}\ (\bibinfo  {publisher}
  {Cambridge university press},\ \bibinfo {year} {2003})\BibitemShut {NoStop}%
\bibitem [{\citenamefont {Zhang}\ and\ \citenamefont
  {Ying}(2002)}]{zhang2002set}%
  \BibitemOpen
  \bibfield  {author} {\bibinfo {author} {\bibfnamefont {S.}~\bibnamefont
  {Zhang}}\ and\ \bibinfo {author} {\bibfnamefont {M.}~\bibnamefont {Ying}},\
  }\href@noop {} {\bibfield  {journal} {\bibinfo  {journal} {Physical Review
  A}\ }\textbf {\bibinfo {volume} {65}},\ \bibinfo {pages} {062322} (\bibinfo
  {year} {2002})}\BibitemShut {NoStop}%
\bibitem [{\citenamefont {Lin}\ \emph {et~al.}(2009)\citenamefont {Lin},
  \citenamefont {Wen}, \citenamefont {Gao},\ and\ \citenamefont
  {Zhu}}]{lin2009eavesdropping}%
  \BibitemOpen
  \bibfield  {author} {\bibinfo {author} {\bibfnamefont {S.}~\bibnamefont
  {Lin}}, \bibinfo {author} {\bibfnamefont {Q.-Y.}\ \bibnamefont {Wen}},
  \bibinfo {author} {\bibfnamefont {F.}~\bibnamefont {Gao}}, \ and\ \bibinfo
  {author} {\bibfnamefont {F.-C.}\ \bibnamefont {Zhu}},\ }\href@noop {}
  {\bibfield  {journal} {\bibinfo  {journal} {Physical Review A}\ }\textbf
  {\bibinfo {volume} {79}},\ \bibinfo {pages} {054303} (\bibinfo {year}
  {2009})}\BibitemShut {NoStop}%
\bibitem [{\citenamefont {Scarani}\ \emph {et~al.}(2004)\citenamefont
  {Scarani}, \citenamefont {Acin}, \citenamefont {Ribordy},\ and\ \citenamefont
  {Gisin}}]{scarani2004quantum}%
  \BibitemOpen
  \bibfield  {author} {\bibinfo {author} {\bibfnamefont {V.}~\bibnamefont
  {Scarani}}, \bibinfo {author} {\bibfnamefont {A.}~\bibnamefont {Acin}},
  \bibinfo {author} {\bibfnamefont {G.}~\bibnamefont {Ribordy}}, \ and\
  \bibinfo {author} {\bibfnamefont {N.}~\bibnamefont {Gisin}},\ }\href@noop {}
  {\bibfield  {journal} {\bibinfo  {journal} {Physical Review Letters}\
  }\textbf {\bibinfo {volume} {92}},\ \bibinfo {pages} {057901} (\bibinfo
  {year} {2004})}\BibitemShut {NoStop}%
\bibitem [{\citenamefont {Fung}\ \emph {et~al.}(2006)\citenamefont {Fung},
  \citenamefont {Tamaki},\ and\ \citenamefont {Lo}}]{fung2006performance}%
  \BibitemOpen
  \bibfield  {author} {\bibinfo {author} {\bibfnamefont {C.-H.~F.}\
  \bibnamefont {Fung}}, \bibinfo {author} {\bibfnamefont {K.}~\bibnamefont
  {Tamaki}}, \ and\ \bibinfo {author} {\bibfnamefont {H.-K.}\ \bibnamefont
  {Lo}},\ }\href@noop {} {\bibfield  {journal} {\bibinfo  {journal} {Physical
  Review A}\ }\textbf {\bibinfo {volume} {73}},\ \bibinfo {pages} {012337}
  (\bibinfo {year} {2006})}\BibitemShut {NoStop}%
\bibitem [{\citenamefont {Zhang}\ \emph {et~al.}(2007)\citenamefont {Zhang},
  \citenamefont {Zou}, \citenamefont {Li}, \citenamefont {Jin},\ and\
  \citenamefont {Guo}}]{zhang2007limitation}%
  \BibitemOpen
  \bibfield  {author} {\bibinfo {author} {\bibfnamefont {S.~L.}\ \bibnamefont
  {Zhang}}, \bibinfo {author} {\bibfnamefont {X.}~\bibnamefont {Zou}}, \bibinfo
  {author} {\bibfnamefont {K.}~\bibnamefont {Li}}, \bibinfo {author}
  {\bibfnamefont {C.}~\bibnamefont {Jin}}, \ and\ \bibinfo {author}
  {\bibfnamefont {G.~C.}\ \bibnamefont {Guo}},\ }\href@noop {} {\bibfield
  {journal} {\bibinfo  {journal} {Physical Review A}\ }\textbf {\bibinfo
  {volume} {76}},\ \bibinfo {pages} {044304} (\bibinfo {year}
  {2007})}\BibitemShut {NoStop}%
\bibitem [{\citenamefont {Li}\ and\ \citenamefont
  {Fang}(2006)}]{jing2006nonorthogonal}%
  \BibitemOpen
  \bibfield  {author} {\bibinfo {author} {\bibfnamefont {J.-B.}\ \bibnamefont
  {Li}}\ and\ \bibinfo {author} {\bibfnamefont {X.-M.}\ \bibnamefont {Fang}},\
  }\href@noop {} {\bibfield  {journal} {\bibinfo  {journal} {Chinese Physics
  Letters}\ }\textbf {\bibinfo {volume} {23}},\ \bibinfo {pages} {775}
  (\bibinfo {year} {2006})}\BibitemShut {NoStop}%
\bibitem [{\citenamefont {Kim}\ and\ \citenamefont
  {Korevaar}(2001)}]{kim2001availability}%
  \BibitemOpen
  \bibfield  {author} {\bibinfo {author} {\bibfnamefont {I.~I.}\ \bibnamefont
  {Kim}}\ and\ \bibinfo {author} {\bibfnamefont {E.~J.}\ \bibnamefont
  {Korevaar}},\ }in\ \href@noop {} {\emph {\bibinfo {booktitle} {Optical
  Wireless Communications IV}}},\ Vol.\ \bibinfo {volume} {4530}\ (\bibinfo
  {organization} {International Society for Optics and Photonics},\ \bibinfo
  {year} {2001})\ pp.\ \bibinfo {pages} {84--95}\BibitemShut {NoStop}%
\bibitem [{\citenamefont {Carrasco-Casado}\ \emph {et~al.}(2014)\citenamefont
  {Carrasco-Casado}, \citenamefont {Denisenko},\ and\ \citenamefont
  {Fernandez}}]{carrasco2014correction}%
  \BibitemOpen
  \bibfield  {author} {\bibinfo {author} {\bibfnamefont {A.}~\bibnamefont
  {Carrasco-Casado}}, \bibinfo {author} {\bibfnamefont {N.}~\bibnamefont
  {Denisenko}}, \ and\ \bibinfo {author} {\bibfnamefont {V.}~\bibnamefont
  {Fernandez}},\ }\href@noop {} {\bibfield  {journal} {\bibinfo  {journal}
  {Optical Engineering}\ }\textbf {\bibinfo {volume} {53}},\ \bibinfo {pages}
  {084112} (\bibinfo {year} {2014})}\BibitemShut {NoStop}%
\bibitem [{\citenamefont {Elmabrok}\ and\ \citenamefont
  {Razavi}(2018)}]{elmabrok2018wireless}%
  \BibitemOpen
  \bibfield  {author} {\bibinfo {author} {\bibfnamefont {O.}~\bibnamefont
  {Elmabrok}}\ and\ \bibinfo {author} {\bibfnamefont {M.}~\bibnamefont
  {Razavi}},\ }\href@noop {} {\bibfield  {journal} {\bibinfo  {journal} {JOSA
  B}\ }\textbf {\bibinfo {volume} {35}},\ \bibinfo {pages} {197} (\bibinfo
  {year} {2018})}\BibitemShut {NoStop}%
\end{thebibliography}%

\end{document}